\documentclass[twocolumn]{aastex63}
\shorttitle{X-RAY EVOLUTION OF SNR 1987A}
\shortauthors{Ravi et al.}
\graphicspath{{./}{figures/}}

\usepackage{tablefootnote}
\usepackage{afterpage}
\usepackage{longtable}
\usepackage{xcolor}
\usepackage{graphics}
\usepackage{epsf}
\usepackage{enumitem}
\usepackage{threeparttable} 

\begin{document}

\title{Latest Evolution of the X-Ray Remnant of SN 1987A: Beyond the Inner Ring}

\correspondingauthor{Aravind Pazhayath Ravi}
\email{apazhayathravi@ucdavis.edu}

\author{Aravind P. Ravi}
\altaffiliation{Current address: Department of Physics and Astronomy, University of California, 1 Shields Avenue, Davis, CA 95616-5270, USA}
\affiliation{Department of Physics, University of Texas at Arlington, Box 19059, Arlington, TX 76019, USA} 

\author{Sangwook Park}
\affiliation{Department of Physics, University of Texas at Arlington,
 Box 19059, 
Arlington, TX 76019, USA} 

\author{Svetozar A. Zhekov}
\affiliation{Institute of Astronomy and National Astronomical Observatory (Bulgarian Academy of Sciences), \\
 72 Tsarigradsko Chaussee Blvd.,
Sofia 1784, Bulgaria}

\author{Salvatore Orlando}
\affiliation{INAF-Osservatorio Astronomico di Palermo,
 Piazza del Parlamento 1,
90134 Palermo, Italy} 

\author{Marco Miceli}
\affiliation{Dipartimento di Fisica e Chimica, Universita degli Studi di Palermo,
 Piazza del Parlamento 1,
90134 Palermo, Italy}
\affiliation{INAF-Osservatorio Astronomico di Palermo,
 Piazza del Parlamento 1,
90134 Palermo, Italy}

\author{Kari A. Frank}
\affiliation{CIERA, Northwestern University,
 1800 Sherman, 8007
Evanston, IL 60201, USA}

\author{Patrick S. Broos}
\affiliation{Department of Astronomy and Astrophysics, Pennsylvania State University, University Park, PA 16802, USA}

\author{David N. Burrows}
\affiliation{Department of Astronomy and Astrophysics, Pennsylvania State University,
University Park, PA 16802, USA}










\begin{abstract}
Based on our Chandra imaging-spectroscopic observations, we present the latest evolution of the X-ray remnant of SN 1987A. Recent changes in the electron temperatures and volume emission measures suggest that the blast wave in SN 1987A is moving out of the dense inner ring structure, also called the equatorial ring (ER). The 0.5--2.0 keV X-ray light curve shows a linearly declining trend (by $\sim$4.5\,\% yr$^{-1}$) between 2016 and 2020, as the blast wave heats the hitherto unknown circumstellar medium (CSM) outside the ER. While the peak X-ray emission in the latest 0.3--8.0 keV image is still within the ER, the radial expansion rate in the 3.0--8.0 keV images suggests an increasing contribution of the X-ray emission from less dense CSM since 2012, at least partly from beyond the ER. It is remarkable that, since 2020, the declining soft X-ray flux has stabilized around $\sim$7 $\times$ 10$^{-12}$ erg s$^{-1}$ cm$^{-2}$, which may signal a contribution from the reverse-shocked outer layers of ejecta as predicted by the 3-D magneto-hydrodynamic (MHD) models. In the latest ACIS spectrum of supernova remnant (SNR) 1987A in 2022 we report a significant detection of the Fe K line at $\sim$6.7 keV, which may be due to changing thermal conditions of the X-ray emitting CSM and/or the onset of reverse shock interactions with the Fe-ejecta.

\end{abstract}

\keywords{supernova remnants, supernovae: 
individual (SNR 1987A) --- X-rays: CSM, ISM }


\section{Introduction} 

SN 1987A is a core-collapse SN discovered on 24 February 1987 in the Large Magellanic Cloud (LMC). It has been the nearest (distance $\sim$51 kpc) and hence the brightest observed SN since the historic Kepler’s SN in 1604 AD \citep[see][for detailed reviews]{Arentt89, McCray93, McCray&Fransson16}. Being the closest observed SN since the advent of modern telescopes, SN 1987A is a unique and crucial astrophysical laboratory to study the early makings of a SNR and a neutron star. Thus, SN 1987A has been observed and studied extensively across the entire electromagnetic spectrum for $>$30 years after the explosion. 

Early optical observations of SN 1987A with the Hubble Space Telescope (HST) revealed an unusual triple-ring structure, the dense CSM that was created by stellar winds from the massive progenitor and then photo-ionized by UV flash from the explosion itself \citep{Burrows95}. These three rings, a bright inner equatorial ring (ER) and two faint outer rings, together form an 
an hourglass structure. The inner ER is within a larger H II region of lower density \citep{Chevalier&Dwarkadas95}. While the origin of these structures is still unclear, it is likely the result of an interaction between the fast moving, low-density blue supergiant wind with the slower, high-density red supergiant wind from different stages of the progenitor’s evolution history \citep{Luo&McCray91, Wang&Mazzali92, Morris_Podsiadlowski07}. The observed X-ray images show a simpler structure with the emission primarily from the shocked ER \citep{Burrows00}. With the best available X-ray spatial and spectral resolutions, the Chandra X-ray Observatory (Chandra) has been crucial to study the photometric, morphological, and spectroscopic evolution of the X-ray remnant of SN 1987A (SNR 1987A, hereafter). 

As part of our on-going Chandra X-ray monitoring program, we have observed SNR 1987A roughly every 6 months for the past 22 years (total of 47 observations as of April 2023). We have presented previous results of our Chandra monitoring campaign of SNR 1987A in the literature \citep[e.g.,][]{Burrows00, Park02, Park04, Park05, Park06, Park11, Zhekov10, Helder13, Frank16}.

Previous results showed that the X-ray emission from SNR 1987A has been dominated by shock interaction with the dense CSM of the ER.  As the shock encountered the main body of the ER around 2004 ($\sim$6200 days after the SN), the expansion rate of the X-ray emitting ring decreased significantly from $\sim$6000 km s$^{-1}$ to $\sim$1700 km s$^{-1}$ \citep{Racusin09}.  Until 2015 ($\sim$10,500 days after the SN), this expansion rate has stayed relatively constant at $\sim$1800 km s$^{-1}$ \citep{Frank16}. The expansion rate of the X-ray emitting ring provides crucial morphological information which can only be resolved by Chandra at these wavelengths.

At the same time as the shock moved into the main body of the dense ER, a sharp upturn was observed in the measured soft X-ray flux in the 0.5--2.0 keV energy range \citep{Park04, Park05}. The X-ray flux continued to increase between 2004 and 2012, suggesting continuous shock-CSM interactions \citep{Park05, Park11, Maggi12, Helder13}. Between 2012 and 2015, the soft X-ray levelled off, hinting at the start of a new evolutionary phase as the blast wave starts to move out of the dense ER \citep{Frank16}. The X-ray light curve is a powerful probe of the density of the shocked CSM when sampled as frequently as our Chandra monitoring program. Both the X-ray light curves and spatially resolved images have been instrumental ingredients for providing observational constraints for modeling the evolution of SN 1987A \citep{Chevalier&Dwarkadas95, Borkowski97a, Zhekov10, Dewey12, Potter14, Orlando15, Orlando19, Orlando20}.

Recent optical, infrared, and radio monitoring observations have suggested that the shock front in SNR 1987A moved out of the dense ER several years ago and is now interacting with the low density CSM that was produced in the red supergiant phase of the progenitor \citep[e.g.,][]{Fransson15, Arendt16, Cendes18, Larsson19, Arendt20}. Our recent deep Chandra gratings spectroscopic study \citep{Ravi21} showed evidence for increasing electron temperatures and decreasing volume emission measure between 2011 and 2018, consistent with expected X-ray emission from the newly shocked CSM.

As the blast wave moves outward, the reverse shock is expected to develop, enclosing the cold ejecta. Three-dimensional MHD simulation models predict that the interaction between the outer layers of the ejecta and the reverse shock would become significant $\gtrsim$35 yr after the SN \citep[i.e., $\gtrsim$2022][]{Orlando15, Orlando20}. In this period, as the blast wave leaves the ER, X-ray emission from the dense ER will fade, while that from the reverse-shocked ejecta will strengthen over time. 

In this paper, we present our latest measurements of X-ray fluxes and the updated Chandra X-ray light curves as well as radial expansion measurements and morphological development of SNR 1987A, covering up to 2023 April (through $\sim$13,200 days after the SN). In Section \ref{sec:2}, we describe our observations and methods of data reduction. In Section \ref{sec:3}, we present X-ray light curves, radial expansion rates, and temporal evolution of the soft X-ray spectrum, including the development of the Fe K line flux. In Section \ref{sec:4}, we discuss physical interpretations of our results, and we summarize them in Section \ref{sec:5}.


\section{Observations and Data Reduction \label{sec:2}} 

The 47 Chandra observations used in this work are presented in Table \ref{Table:Observations}. Results from our Chandra monitoring observations between 1999 and 2015 have been previously discussed in the literature \citep[]{Burrows00, Park02, Park04, Park05, Park06, Racusin09, Park11, Helder13, Frank16}. Since 2016, there have been 15 new observations as part of our monitoring campaign (eight ACIS-S/HETG and seven HRC-S/LETG observations). Our Chandra monitoring observation taken in September 2022 (ObsID 27443 with the HRC-S/LETG configuration) was performed with a significantly shorter exposure time (13 ks) than our scheduled duration (59 ks) due to an unexpected anomaly that occurred in the detector. Thus, the utility of this observation was limited, and we do not include this data set in our analysis.  

As discussed in \cite{Frank16} and \cite{Helder13}, the ACIS observing configurations have been changed multiple times since 2008 to mitigate the photon pileup effects due to the very high count rates from the brightening supernova remnant. Since 2008, HETG has been inserted to further reduce the pileup effects of the ACIS detector. In the epochs of Jul 2008 and Jan 2009, we observed SNR 1987A both with and without HETG for calibration purposes to ensure a smooth transition between the two configurations \citep{Frank16}. For these epochs, we use the bare ACIS data for imaging due to better count statistics. For spectral analysis at these epochs, we use the HETG observations, as the use of the grating reduces pileup in the observed CCD spectrum. Additionally, the varying molecular contamination on the optical blocking filter \citep[OBF; see][]{O'Dell13, Plucinsky18} of the ACIS detector leads to incorrect photon counts from standard data reduction methods, which needs to be accounted for. We added the HRC-S/LETG configuration since September 2015 for flux calibration purposes, as the HRC detector is not affected by this contamination.

We re-processed and analyzed all previously-published data taken between 1999 and 2015, and include them in this work for a self-consistent analysis of the entire data set. Following our approaches in \cite{Helder13} and \cite{Frank16}, we exclude the earliest two observations, ObsIDs 1387 and 122 in our spectral analysis, as the focal plane temperature for those observations was higher than normal, and we have no corresponding charge transfer inefficiency correction available \footnote{https://cxc.cfa.harvard.edu/ciao/why/cti.html}. Additionally, they have an order of magnitude lower counts in the 0.5--8.0 keV band. Only imaging data are used at these epochs.

\begin{longtable*}{ccccccccc}
\caption{\mbox{SNR 1987A: Chandra Observations, Fluxes, and Radii}} \label{Table:Observations} \\
\hline \multicolumn{1}{c}{Epoch} & \multicolumn{1}{c}{Age} & \multicolumn{1}{c}{ObsID$^{a}$} & \multicolumn{1}{c}{Instrument} & \multicolumn{1}{c}{Exposure$^{b}$} & \multicolumn{1}{c}{Counts$^{c}$} & \multicolumn{1}{c}{0.5--2.0 keV Flux$^{d}$} & \multicolumn{1}{c}{3.0--8.0 keV Flux$^{e}$} &  \multicolumn{1}{c}{Radius$^{f}$} \\ \hline 
\endfirsthead



\hline \hline
\multicolumn{9}{l}%
{{90\,\% confidence intervals and 1-$\sigma$ errors are the quoted uncertainties for flux and radii measurements, respectively.}}\\
\multicolumn{9}{l}%
{{Horizontal line after 2015 Sep separates the published and new observations used in this work.}}\\
\multicolumn{9}{l}%
{{$^\mathrm{a}$Only the longest ObsID in every epoch shown.}}\\
\multicolumn{9}{l}%
{{$^\mathrm{b}$Total exposure of all ObsIDs in an epoch used for flux measurement.}}\\
\multicolumn{9}{l}%
{{$^\mathrm{c}$Counts in the ACIS 0th-order spectra and LETG 1st-order (+1/-1) spectra in 0.5--8.0 keV energy band.}}\\
\multicolumn{9}{l}%
{{$^\mathrm{d,}$$^\mathrm{e}$X-ray fluxes in units of 10$^{-13}$ erg s$^{-1}$ cm$^{-2}$.}}\\
\multicolumn{9}{l}%
{{$^\mathrm{f}$X-ray radius (in unts of arcsec) measured only from the longest ObsID in the epoch.}}\\
\multicolumn{9}{l}%
{{$^\mathrm{g}$ACIS observations with a Science Instrument Module (SIM) offset of $\sim$8 mm along -Z axis.}}\\
\endlastfoot

1999 Oct & 4608 & 1387  & ACIS-S/HETG & 68.9  & 342 & \ldots & \ldots & 0.575 $\pm$ 0.024 \\
2000 Jan & 4711 & 122 & ACIS & 8.6 & 530 & \ldots & \ldots & 0.629 $\pm$ 0.022\\
2000 Dec & 5036 & 1967 & ACIS & 98.8 & 8721 & 2.76$^{+0.05}_{-0.04}$ & 0.74$^{+0.06}_{-0.07}$ & 0.666 $\pm$ 0.007 \\
2001 Apr & 5175 & 1044 & ACIS & 17.8 & 1680 & 3.04$^{+0.08}_{-0.11}$ & 1.04$^{+0.22}_{-0.25}$ & 0.698 $\pm$ 0.015\\
2001 Dec & 5406 & 2831 & ACIS & 49.4 & 5989 & 4.20$^{+0.06}_{-0.11}$ & 0.96$^{+0.12}_{-0.13}$ & 0.701 $\pm$ 0.008\\
2002 May & 5560 & 2832 & ACIS & 44.3 & 6204& 5.20$^{+0.07}_{-0.12}$ & 1.16$^{+0.16}_{-0.19}$ & 0.711 $\pm$ 0.008\\
2002 Dec & 5789 & 3829 & ACIS & 49.0 & 8514& 7.37$^{+0.11}_{-0.10}$ & 1.30$^{+0.11}_{-0.14}$ & 0.730 $\pm$ 0.007\\
2003 Jul & 5978 & 3830 & ACIS & 45.3 & 9354& 9.42$^{+0.12}_{-0.13}$ & 1.57$^{+0.14}_{-0.20}$ & 0.746 $\pm$ 0.007\\
2004 Jan & 6157 & 4614 & ACIS & 46.5 & 11527& 11.57$^{+0.12}_{-0.12}$ & 1.88$^{+0.16}_{-0.31}$ & 0.754 $\pm$ 0.006\\
2004 Jul & 6359 & 4615 & ACIS & 48.8 & 17525& 14.78$^{+0.14}_{-0.17}$ & 1.99$^{+0.17}_{-0.17}$ & 0.753 $\pm$ 0.005\\
2005 Jan & 6530 & 5579 & ACIS & 31.9 & 15864& 17.68$^{+0.16}_{-0.21}$ & 2.07$^{+0.19}_{-0.18}$ & 0.746 $\pm$ 0.005\\
2005 Jul & 6713 & 5580 & ACIS & 23.2 & 13843& 21.93$^{+0.21}_{-0.29}$ & 2.67$^{+0.32}_{-0.39}$ & 0.758 $\pm$ 0.006\\
2006 Jan & 6913 & 6668 & ACIS & 42.3 & 30161 & 26.87$^{+0.16}_{-0.24}$ & 3.25$^{+0.22}_{-0.25}$ & 0.755 $\pm$ 0.004\\
2006 Jul & 7094 & 6669 & ACIS & 42.3 & 29718 & 31.91$^{+0.10}_{-0.23}$ & 3.54$^{+0.22}_{-0.19}$& 0.766 $\pm$ 0.004\\
2007 Jan & 7270 & 7636 & ACIS & 33.5 & 31990& 38.25$^{+0.28}_{-0.32}$ & 4.02$^{+0.30}_{-0.27}$ & 0.770 $\pm$ 0.004\\
2007 Jul & 7445 & 7637 & ACIS & 23.4 & 25637 & 43.02$^{+0.49}_{-0.40}$ & 3.66$^{+0.23}_{-0.35}$ & 0.780 $\pm$ 0.004\\
2008 Jan & 7624 & 9142 & ACIS & 6.4 & 7958 & 47.22$^{+0.76}_{-0.84}$ & 4.09$^{+0.69}_{-0.98}$& 0.777 $\pm$ 0.007\\
2008 Jul & 7799  & 9144 & ACIS-S/HETG & 42.0 &  5049& 52.39$^{+1.22}_{-1.89}$ & 5.02$^{+0.29}_{-0.68}$ & \ldots \\
2008 Jul & 7802 & 9143 & ACIS & 8.6 & 11840 & \ldots  & \ldots  & 0.777 $\pm$ 0.006 \\
2009 Jan & 7987 & 10130 & ACIS & 6.0 & 8913& \ldots & \ldots & 0.787 $\pm$ 0.007\\
2009 Jan & 8000 & 10855 & ACIS-S/HETG & 18.8 & 2415 & 57.20$^{+2.13}_{-1.94}$ & 5.40$^{+0.64}_{-0.54}$ & \ldots\\
2009 Jul & 8169 & 10222 & ACIS-S/HETG & 24.4 & 3450 & 63.47$^{+2.06}_{-1.67}$ & 6.49$^{+0.55}_{-1.52}$ & 0.776 $\pm$ 0.011\\ 
2010 Mar & 8433 & 11090$^\mathrm{g}$ & ACIS-S/HETG & 24.6 & 3503 & 65.16$^{+2.02}_{-1.95}$ & 6.64$^{+0.57}_{-0.53}$ & 0.780 $\pm$ 0.011\\ 
2010 Sep & 8617 & 13131$^\mathrm{g}$ & ACIS-S/HETG & 26.5 & 3954 & 67.76$^{+1.94}_{-2.49}$ & 6.91$^{+0.68}_{-1.15}$ & 0.785 $\pm$ 0.011\\ 
2011 Mar & 8796 & 12539$^\mathrm{g}$ & ACIS-S/HETG & 52.2 &  8216 & 70.97$^{+1.24}_{-1.34}$ & 8.16$^{+0.38}_{-0.34}$& 0.779 $\pm$ 0.007\\ 
2011 Sep & 8975 & 12540$^\mathrm{g}$ & ACIS-S/HETG & 37.5 &  6133 & 74.36$^{+1.22}_{-1.83}$ & 8.16$^{+0.37}_{-0.38}$ & 0.779 $\pm$ 0.009\\ 
2012 Mar & 9165 & 13735$^\mathrm{g}$ & ACIS-S/HETG & 42.9 &  7289 & 81.48$^{+1.66}_{-1.83}$ & 8.97$^{+0.69}_{-0.70}$ & 0.793 $\pm$ 0.008\\ 
2013 Mar & 9523 & 14697 & ACIS-S/HETG & 67.2 & 11674 & 78.73$^{+1.53}_{-1.50}$ & 10.24$^{+0.49}_{-0.40}$ & 0.809 $\pm$ 0.007\\ 
2013 Sep & 9713 & 14698 & ACIS-S/HETG & 68.5 & 11998 & 82.56$^{+1.45}_{-1.24}$ & 11.50$^{+0.29}_{-0.36}$ & 0.807 $\pm$ 0.006\\ 
2014 Mar & 9885 & 15809 & ACIS-S/HETG & 70.5 & 12051 & 80.42$^{+1.36}_{-1.30}$ & 12.07$^{+0.30}_{-0.38}$ & 0.813 $\pm$ 0.006\\ 
2014 Sep & 10071 & 15810 & ACIS-S/HETG & 47.9 & 8048 & 83.38$^{+1.68}_{-1.53}$ & 11.83$^{+0.60}_{-0.82}$ & 0.826 $\pm$ 0.008\\ 
2015 Mar & 10246 & 16757 & HRC-S/LETG & 67.6 & 6898& 82.58$^{+3.33}_{-3.36}$ & \ldots & 0.755 $\pm$ 0.009 \\
2015 Sep & 10433 & 16756 & ACIS-S/HETG & 66.3 & 10500 & 82.52$^{+1.64}_{-1.49}$ & 12.72$^{+0.59}_{-0.13}$ & 0.828 $\pm$ 0.007  \\
\hline 
2016 Mar & 10626 & 17898 & HRC-S/LETG & 68.6 & 8324& 83.03$^{+3.18}_{-3.79}$ & \ldots & 0.767 $\pm$ 0.014 \\
2016 Sep & 10805 & 19882 & ACIS-S/HETG & 66.9 & 10150& 85.20$^{+1.67}_{-1.97}$ & 14.52$^{+0.57}_{-0.83}$ & 0.835 $\pm$ 0.009  \\
2017 Mar & 10979 & 19290& HRC-S/LETG & 67.6 & 8119 & 82.01$^{+3.98}_{-3.12}$ & \ldots & 0.775 $\pm$ 0.011 \\
2017 Sep & 11168 & 20793& ACIS-S/HETG & 66.9 & 9189 & 78.30$^{+2.23}_{-2.57}$ & 14.18$^{+0.46}_{-0.76}$ & 0.833 $\pm$ 0.009  \\
2018 Mar & 11351 & 21042 & ACIS-S/HETG & 313 & 39783 & 74.79$^{+0.90}_{-0.92}$ & 15.27$^{+0.25}_{-0.23}$ & 0.847 $\pm$ 0.010  \\
2018 Sep & 11527 & 20277 & ACIS-S/HETG & 67.2 & 8858 & 75.58$^{+2.23}_{-2.32}$ & 14.95$^{+0.65}_{-0.44}$ & 0.850 $\pm$ 0.011  \\
2019 Mar & 11716 & 22155 & HRC-S/LETG & 77.6 & 9233 & 74.68$^{+3.93}_{-3.45}$ & \ldots & 0.779 $\pm$ 0.014 \\
2019 Sep & 11849 & 21304 & ACIS-S/HETG & 81.7 & 9682& 72.43$^{+2.40}_{-2.25}$ & 16.55$^{+0.73}_{-0.51}$ & 0.861 $\pm$ 0.011  \\
2020 Mar & 12087 & 23198 & HRC-S/LETG & 84.6 & 10173 & 71.70$^{+3.32}_{-2.97}$ & \ldots & 0.784 $\pm$ 0.011 \\
2020 Sep & 12255 & 22425& ACIS-S/HETG & 89.9 & 10209 & 70.68$^{+1.85}_{-1.78}$ & 17.92$^{+0.58}_{-0.43}$ & 0.862 $\pm$ 0.009  \\
2021 Mar & 12440 & 24653 & HRC-S/LETG & 58.6 & 6811 & 69.26$^{+3.94}_{-4.28}$ & \ldots & 0.786 $\pm$ 0.013 \\
2021 Oct & 12666 & 24654 & ACIS-S/HETG & 87.7 & 9299 & 69.51$^{+1.85}_{-1.78}$ & 16.86$^{+0.54}_{-0.52}$ & 0.868 $\pm$ 0.014  \\
2022 Sep & 12995 & 25514 & ACIS-S/HETG & 83.4 & 8595 & 69.44$^{+2.07}_{-1.91}$ & 17.74$^{+0.52}_{-0.58}$ & 0.871 $\pm$ 0.010  \\
2023 Apr & 13197  & 27495 & HRC-S/LETG & 40.9 & 4701 & 66.16$^{+5.75}_{-4.48}$ & \ldots & 0.814 $\pm$ 0.022\\
\end{longtable*}

In this work, we analyze the 0th-order images and spectra from ACIS-S/HETG observations to measure the evolving X-ray radii and fluxes, respectively. We use the 0th-order HRC-S/LETG images and the 1st-order dispersed spectra (in the 0.5--2 keV band) to cross-check the residual pileup effects in the 0th-order ACIS data.

For the data reduction, we adopt a similar procedure to that presented in our previous works \citep[e.g.,][]{Burrows00, Racusin09, Park11, Helder13, Frank16}. We processed our entire data set between 2000 and 2023 with the \verb|chandra_repro| script within CIAO version 4.15. We use the calibration database CALDB version 4.10.2 (with the latest ACIS-OBF molecular contamination models) for our spectral analysis. For each ACIS observation, we extracted the 0th-order spectrum using the CIAO script \verb|specextract| from a circular region with a radius of $\sim$4$^{\prime \prime}$ centered on the source. We extracted the background spectrum from an annular region with an inner radius of $\sim$6$^{\prime \prime}$ and an outer radius of $\sim$12$^{\prime \prime}$. 

Since 2016, all of our Chandra observations of SNR 1987A at each epoch were split into multiple segments for effective telescope operations. Thus, for our spectral analysis, we combined spectra of all individual ObsIDs for a given epoch. While the photon pileup is significantly reduced in the 0th-order spectra thanks to the insertion of HETG, there remain some marginal pileup effects (typically with $\sim$\,5\,\% of the total counts in the observed spectrum between 0.5--8.0 keV being affected by pileup\footnote{We choose not to use the terms “pileup fraction” or “pileup percentage” for defining this metric because the ACIS community has several conflicting definitions for those terms; see Section 1.2 in “The Chandra ABC Guide to Pileup” (http://cxc.harvard.edu/ciao/download/doc/pileup\_abc.pdf).}). To further combat these residual pileup effects, we used the ACIS pileup simulator \citep{Broos10} to synthesize the pileup-corrected ACIS spectra for each epoch, following the approach as described in \cite{Helder13}. From the HRC-S/LETG data, we extracted 1st-order dispersed spectra (positive and negative arms) with the CIAO script, \verb|tgextract|. Both the 0th-order spectra (from ACIS observations) and the 1st-order LETG spectra (from HRC-S/LETG observations) were re-binned to have at least 30 counts per energy channel. For our spectral analysis we use the version 12.13.0c of the XSPEC software package \citep{Arnaud96}.

We performed our imaging analysis generally following methods that were used in our previous works \citep[e.g.,][]{Burrows00, Racusin09, Park11, Helder13, Frank16}. Considering the Chandra astrometric uncertainties for the ACIS data of SNR 1987A \citep[$\sim$0\farcs 1 - e.g.,][]{Park02} and its angular size ($\sim$1\farcs 6), directly combining Chandra images of SNR 1987A from multiple ObsIDs at each observation epoch is difficult. Thus, we used only the longest ObsID in each observation for our image analysis. The standard ACIS spatial resolution is $\sim$0\farcs 5, but SNR 1987A has an angular size of $\sim$1\farcs 6, thus, it is barely resolved by the ACIS detector. For our imaging analysis, we apply the sub-pixel resolution and point spread function (PSF)-deconvolution methods to improve the effective ACIS resolution, which has become a standard image processing method for the Chandra analysis of SNR 1987A \citep[e.g.,][]{Park11, Helder13, Frank16}. The subpixel resolution based on split-pixel events (which is now part of the standard Chandra data processing with CIAO) may improve the effective ACIS resolution by $\sim$10\,\% \citep{Mori01}. We generated three sub-band images of sub-pixel resolution, with energy ranges of 0.3--8.0 keV (broad), 0.5--2.0 keV (soft), and 3.0--8.0 keV (hard) for each ACIS observation.

For the PSF deconvolution, we binned the raw ACIS images to the 1/4 pixel size in both X and Y dimensions to create a 158 $\times$ 158 pixel image. We performed Chandra Ray Tracer (ChaRT) simulations to obtain the best available PSF, assuming monochromatic energies of 0.92 keV, 2.3 keV, and 3.8 keV for soft, broad, and hard ACIS sub-band images, respectively. 

HRC images have no energy resolution and smaller pixel scales than ACIS images (about 1/4 of the ACIS pixel size).  We binned the HRC images to 0.954 pixel size in both X and Y dimensions of the detector to create a similarly sized 158 $\times$ 158 pixel image. We also used ChaRT simulations to construct the best available HRC PSFs assuming a monochromatic energy of 1.5 keV, approximately corresponding to the mean of the photon energy band covered by the HRC.

For each ObsID, the simulated PSF was generated by combining the results of 10 iterations with the web-based ChaRT interface\footnote{https://cxc.cfa.harvard.edu/ciao/PSFs/chart2/runchart.html}, each done with a different random seed to reduce the overall noise in the PSF. All re-binned images were then deconvolved with the simulated PSFs using the Lucy-Richardson iterative algorithm \citep{Richardson72, Lucy74} and smoothed by convolving with a Gaussian (FWHM $\sim$0\farcs 1). The effective spatial resolution of the ACIS images after the PSF deconvolution is $\sim$0\farcs 27, which is in agreement with the results presented in previous works \citep[e.g.,][]{Park11, Helder13, Frank16}. 

\section{Analysis and Results} \label{sec:3}

\subsection{Broadband Spectral Model Fits} \label{sec:3.1}
\setcounter{figure}{0}
\begin{figure*} 
     \hspace{-0.2cm}
\begin{longtable*}{cc}
\includegraphics[scale=0.35,keepaspectratio]{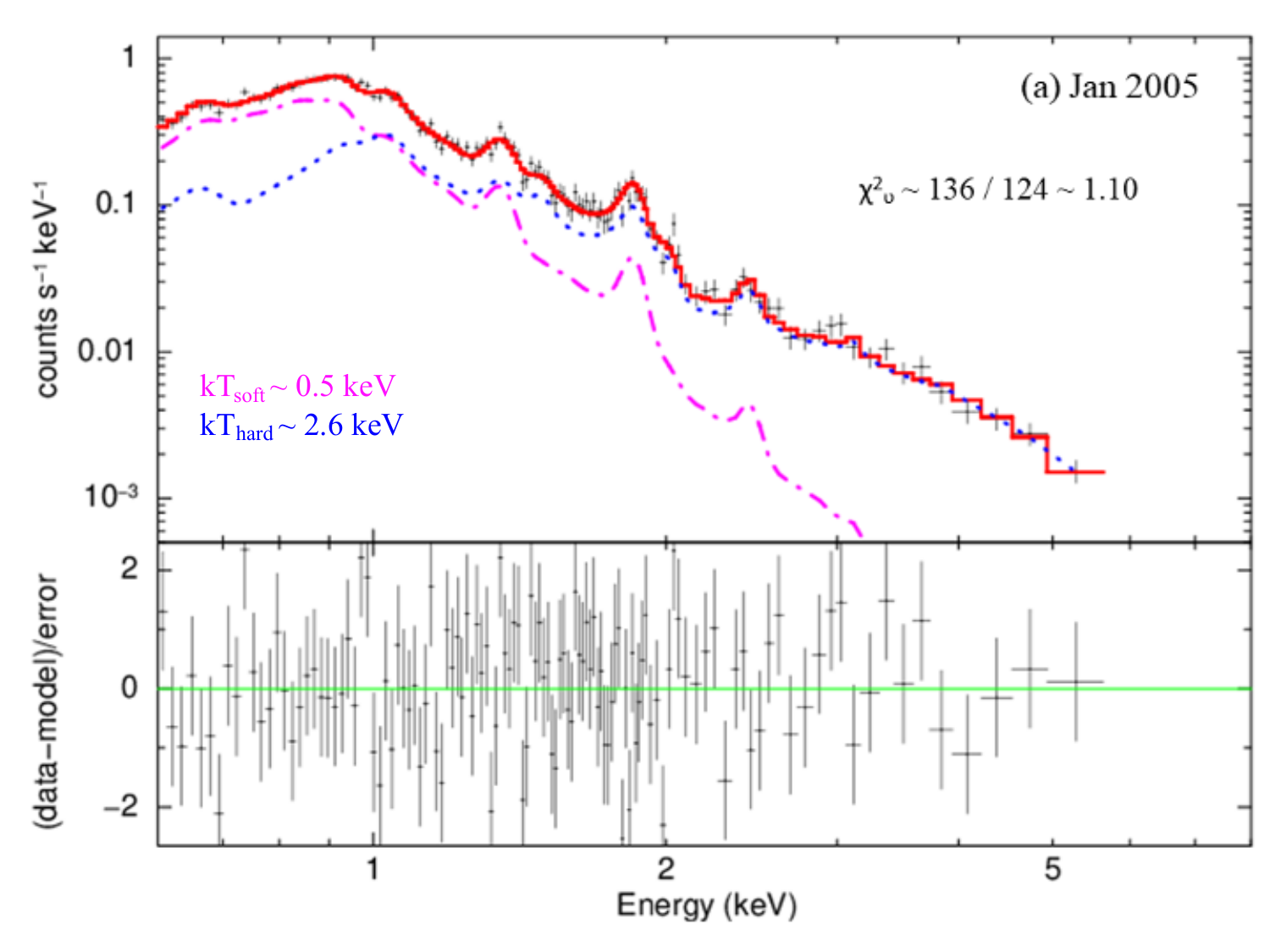} & \includegraphics[scale=0.35,keepaspectratio]{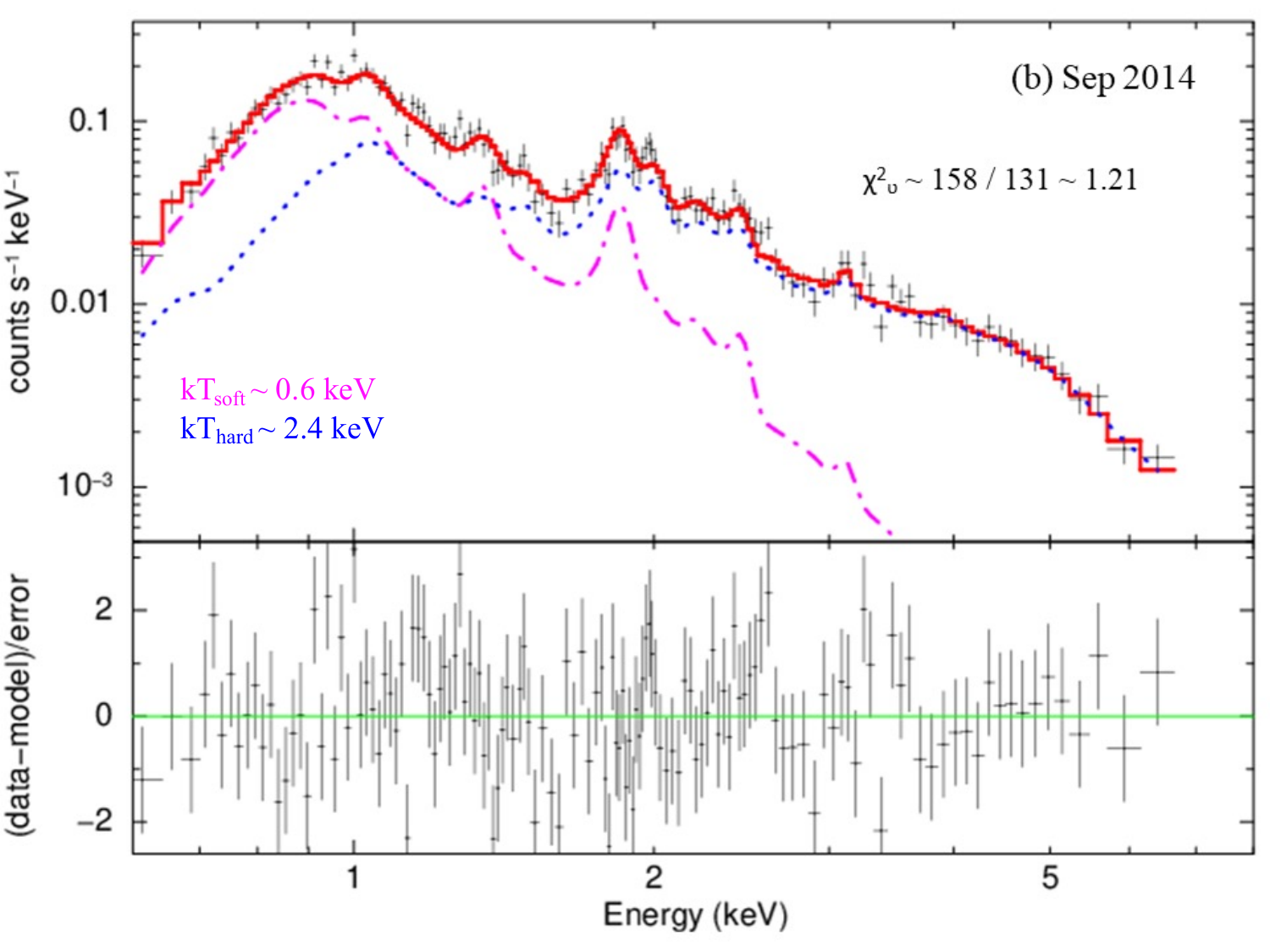} \\
\includegraphics[scale=0.35,keepaspectratio]{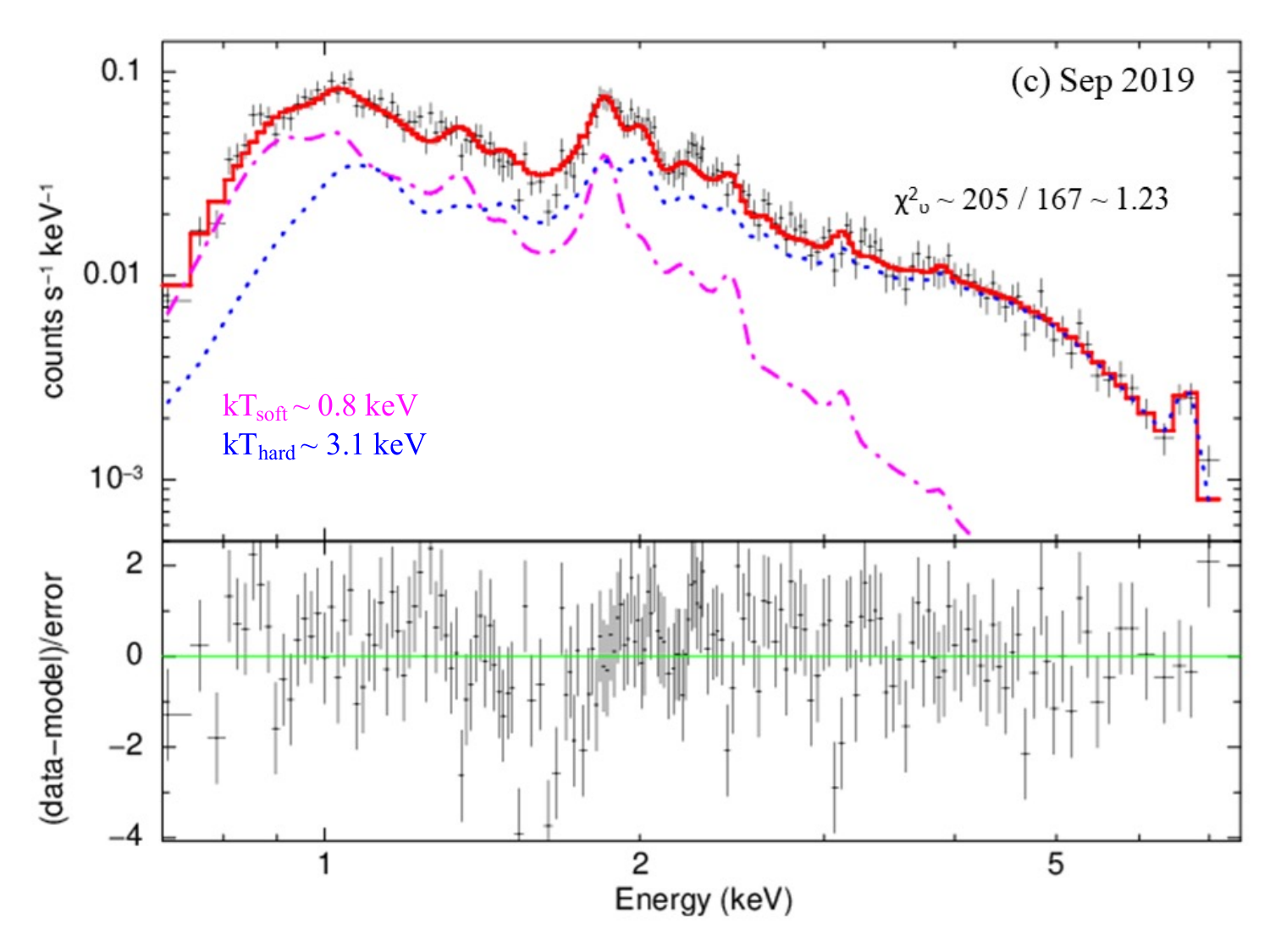} &
\includegraphics[scale=0.35,keepaspectratio]{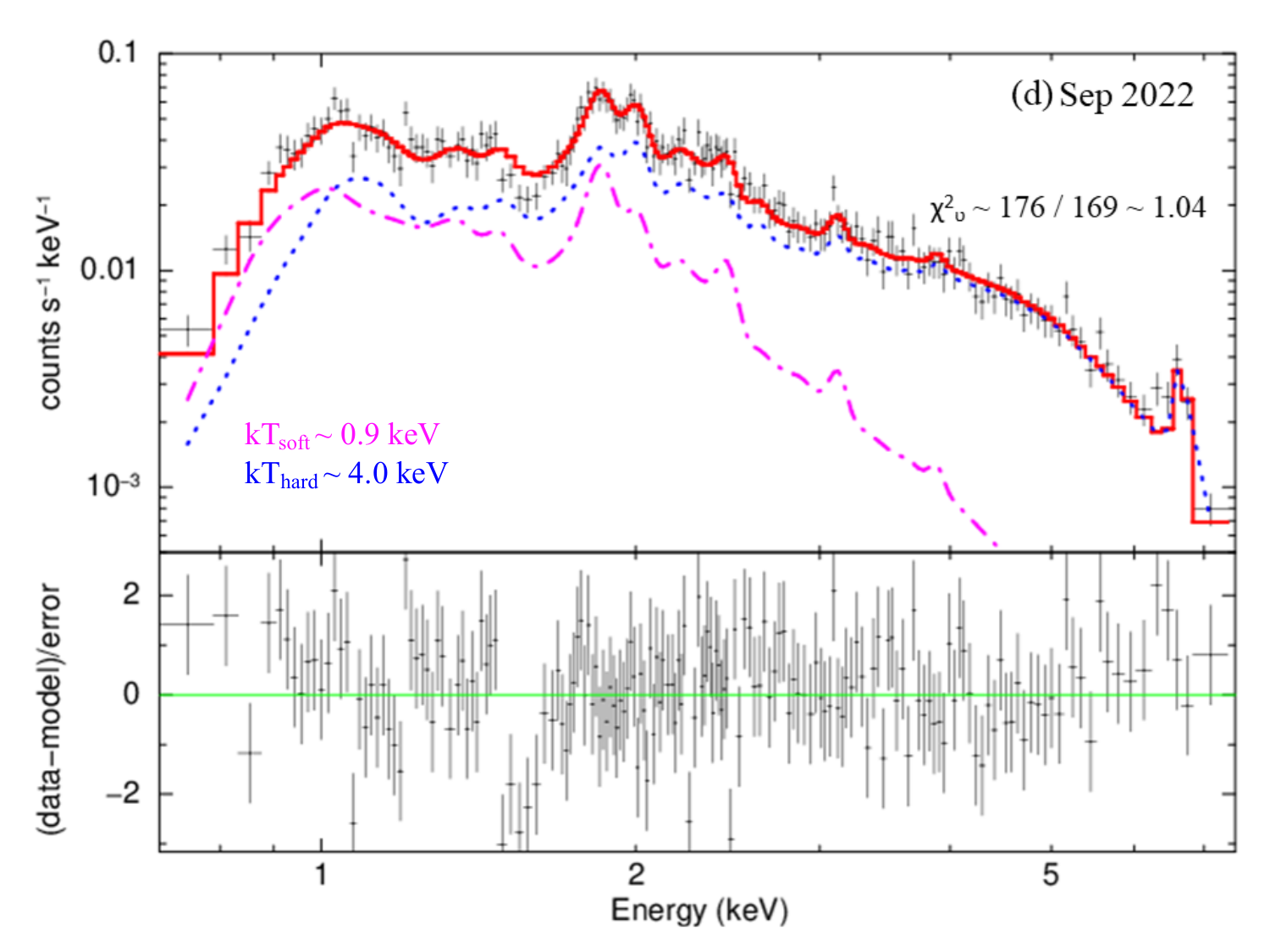} \\
\end{longtable*}
\caption{Best-fit two-component models (in red) and observed ACIS-S / HETG 0th-order combined spectra (in black) in (a) January 2005, (b) September 2014, (c) September 2019, and (d) September 2022. The soft and hard model components are shown as dot-dash (in magenta) and dotted (in blue) lines, respectively. Post-shock electron temperatures associated with the soft and hard components are marked. The reduced $\chi^{2}_{\nu}$ value associated with the best-fit model is shown in each panel.}
\label{Fig:spectral fits}
\end{figure*}
We perform broadband spectral model fits for the 0th-order ACIS-S/HETG (``ACIS spectrum'' hereafter) and the 1st-order HRC-S/LETG (``LETG spectrum'', hereafter) spectra listed in Table \ref{Table:Observations}. Based on previous analyses of the ACIS spectra, it has been shown that spectral model fits with two characteristic shock components are required to adequately describe the X-ray emitting plasma in SNR 1987A \citep[e.g.,][]{Park05, Park06, Park11, Helder13, Frank16}. X-ray emitting plasma in SNR 1987A has been shown to have significant non-equilibrium effects from the high-resolution Chandra gratings spectral analyses \citep{Zhekov06, Dewey08, Zhekov09, Bray20, Ravi21}. Thus, we fitted the ACIS and LETG spectra with a two-component plane-parallel shock model (\verb|vpshock| in XSPEC) that takes into account the NEI effects in a optically-thin hot plasma \citep{Borkowski01}. 

We note that the early observation in 2001 April (ObsID: 1044) has significantly lower number of counts (by several times or more compared to those at other epochs; Table \ref{Table:Observations}). Due to these poor count statistics, a single-component spectral model was sufficient to fit the broadband spectrum with an averaged gas temperature \citep[e.g.,][]{Park11, Helder13, Frank16}. In our spectral model fits, we implement the NEI version 3.0 in XSPEC, based on atomic database values in AtomDB v3.0.9 \footnote{http://atomdb.org/}. We show the observed ACIS spectra with our best-fit two-component shock model at several representative epochs in Figure \ref{Fig:spectral fits}. 
\begin{figure*} 
     \centering
     \hspace{-0.2cm}
\begin{longtable*}{cc}
\includegraphics[scale=0.35,keepaspectratio]{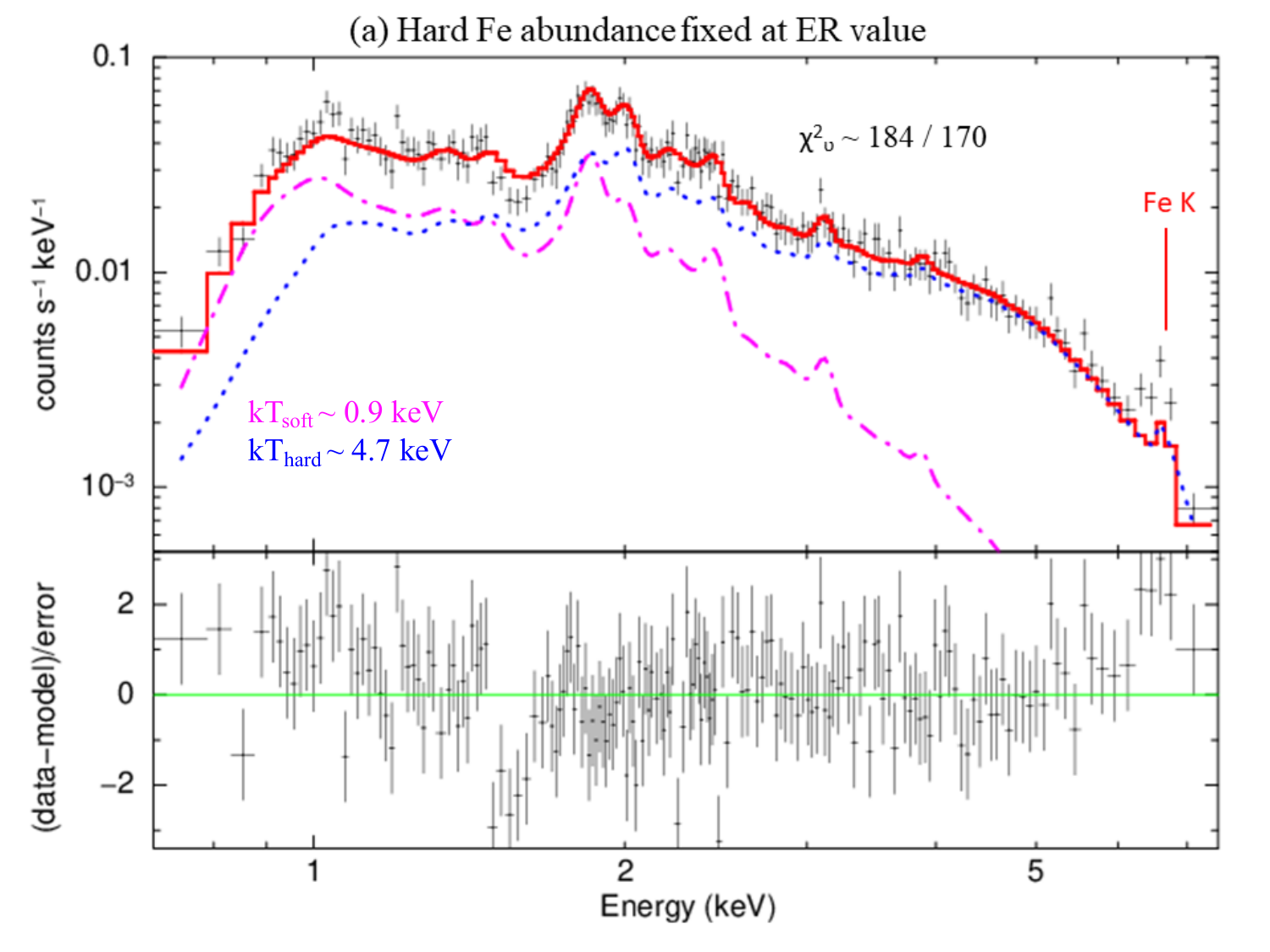} & \includegraphics[scale=0.35,keepaspectratio]{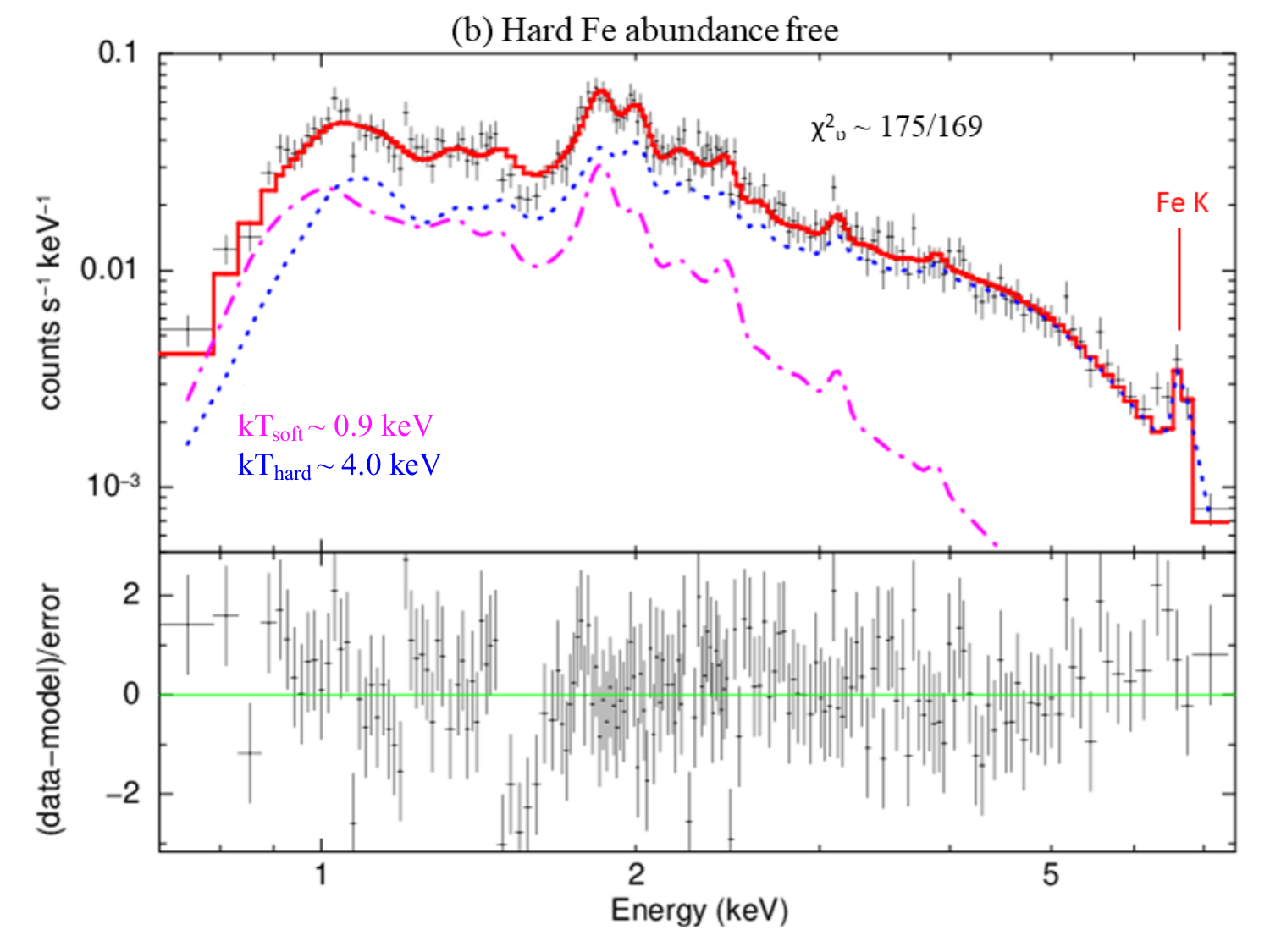} \\
\end{longtable*}
\caption{Best-fit two-component model (in red) and observed 0th-order spectrum (in black) from the latest ACIS-S/HETG observation in September 2022 (ObsID: 25514). Plot schemes for soft and hard components are the same as Figure \ref{Fig:spectral fits}.  In panel (a), all abundances for both soft and hard components are fixed at ER values.  The Fe K line complex at $\sim$6.7 keV (marked in red) is underestimated in the left panel. In panel (b), the Fe abundance associated with the hard component is allowed to vary while all other abundances are fixed at ER values.}
\label{Fig:ER abund compared}
\end{figure*}

For the LETG spectra, we fit the 1st-order dispersed spectrum (both of the positively- and negatively-dispersed spectra), simultaneously. To model the higher-resolution emission lines in the grating spectra we convolved the two-component shock models with a Gaussian smoothing model (\verb|gsmooth|) as demonstrated previously in the literature \citep[e.g.,][]{Zhekov06, Zhekov09}.

In our spectral model fits, we allowed the electron temperatures ($kT$, where where the Boltzmann constant $k =$ 1.38 $\times$ 10$^{-23}$ J K$^{-1}$ and $T$ is the temperature in K), ionization ages ($\tau = n_{e}t$, where $n_{e}$ is the electron density of the shocked gas and $t$ is the time since the gas was shocked), and normalizations of both components to vary. Hereafter, the lower $kT$ among the two characteristic model components is termed the soft component and the higher $kT$ is the hard component. 

In spectral model fits of the ACIS data taken in September 2020, we found that the best-fit electron temperature for the soft component ($kT_\mathrm{soft}$ = 0.45$^{+0.21}_{-0.16}$ keV) was abruptly lower than those estimated in the last several years (typically $kT_\mathrm{soft}$ $\sim$ 0.7--0.9 keV) but with larger uncertainties. We repeated our spectral model fits for this particular data set with $kT_\mathrm{soft}$ fixed at 0.78 keV, the best-fit electron temperature that we measured for our 1st-order LETG spectrum taken in March 2020, the closest epoch to September 2020 among our Chandra monitoring observations of SNR 1987A. We found that these two spectral model fits ($\chi^{2}_{\nu}$: 196/182 v/s 200/183) are statistically indistinguishable (F-test probability $\sim$0.05) and thus consider it as a statistical fluctuation. In this work, we adopt the results from the latter fits (with $kT_\mathrm{soft}$ = 0.78 keV), because we concluded that the abrupt change of $kT_\mathrm{soft}$ $\sim$0.45 keV only in the single epoch of September 2020 may be physically unrealistic. Nonetheless, the best-fit model with $kT_\mathrm{soft}$ = 0.45 keV involves large uncertainties, and thus, our scientific outcomes are not affected by accepting either result.

The total absorption column in the direction of SNR 1987A was fixed at $N_\mathrm{H}$ = 2.17 $\times$ 10$^{21}$ cm$^{-2}$ \citep{Ravi21}.  All abundances are relative to the solar abundance table by \cite{Asplund09} - \verb|aspl| in XSPEC. Initially, we fixed the following elemental abundances for SNR 1987A: He = 1.98 \citep{Matilla10}, C = 0.12 \citep{Fransson&Lunqvist96}, N = 0.92, O = 0.14 \citep{Zhekov09}, Ne = 0.34, Mg = 0.25, Si = 0.36, S = 0.40, and Fe = 0.19 \citep{Ravi21}. Further, we also fix Ar = 0.776, Ca = 0.354, and Ni = 0.662 at their respective LMC values \citep{Russell&Dopita92}. We tie all the elemental abundances between the soft and hard components. With these spectral model fits, we are able to describe the observed spectra that we obtained until $\sim$2016. These results are in agreement with the previously published results in \cite{Frank16}.  

\begin{figure} 
     \hspace{-0.2cm}
\includegraphics[width=0.5\textwidth,keepaspectratio]{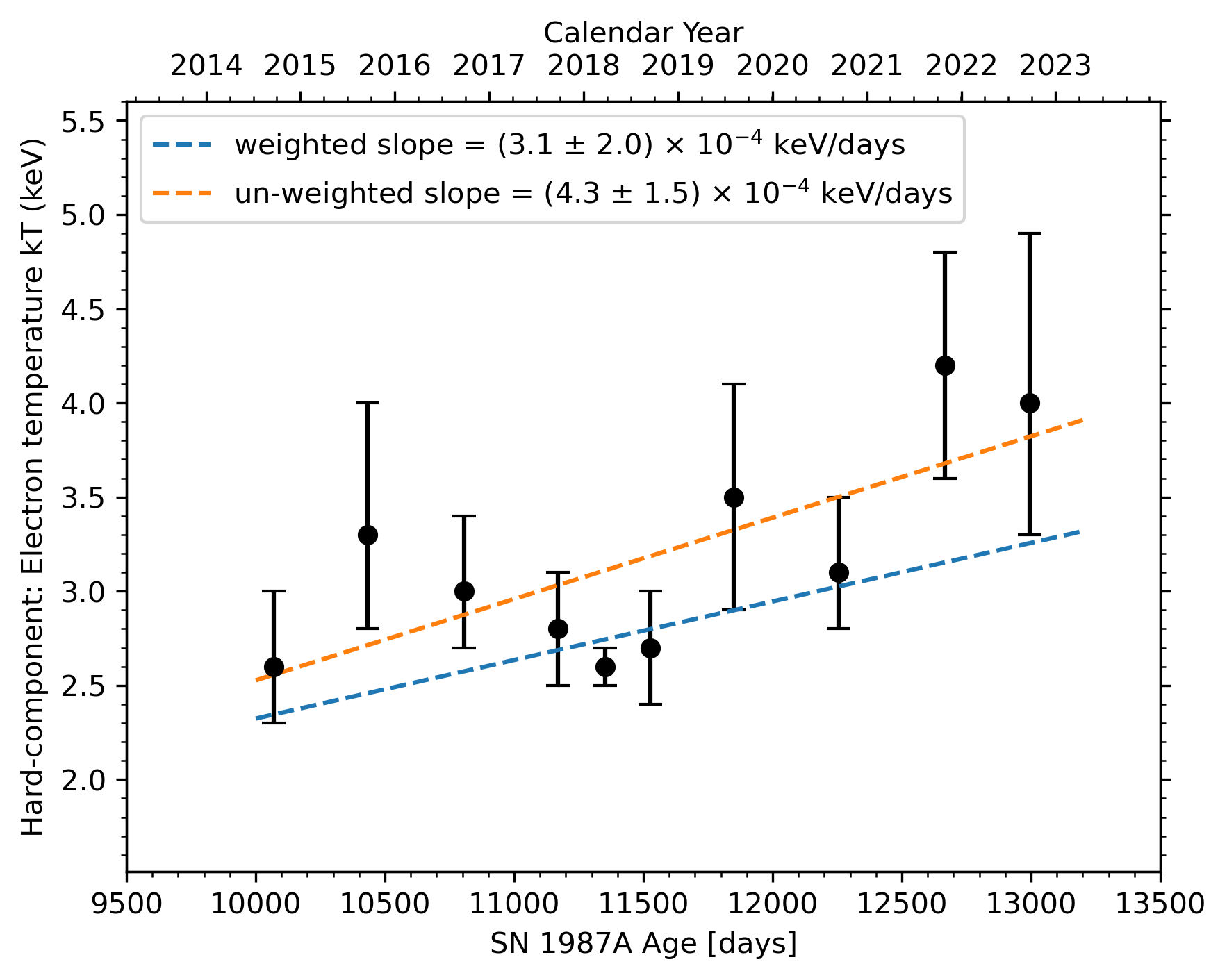} 
\caption{Measured electron temperatures associated with the hard component between September 2016 and September 2022. The best-fit line considering weighted data (slope = (3.1 $\pm$ 2.0) $\times$ 10$^{-4}$ keV/day) is shown with a dashed blue line while the linear fit based on unweighted data (slope = (4.3 $\pm$ 1.5) $\times$ 10$^{-4}$ keV/day) is shown with a dashed orange line. A marginal linear increase with the weighted data} suggests that the increasingly significant line around $\sim$6.7 keV is likely due to the He-like triplet Fe XXV.
\label{Fig:kT hot}
\end{figure}

Starting in March 2018, we detect emission from a clear line-like feature at $\sim$6.7 keV, likely due to the presence of the Fe K line complex. The detection of Fe K emission line features has been reported in the XMM-Newton (XMM) data since $\sim$2009  \citep[e.g.,][]{Strum10, Maggi12, Sun_2021}, and here we confirm it with our Chandra data. We find that fixing the Fe abundance at the ER value (0.19) during these epochs significantly underestimates the observed fluxes at $\sim$6.7 keV. In Figure \ref{Fig:ER abund compared}a, we show the ACIS spectrum taken in September 2022 overlaid with the spectral model fits with the Fe abundances fixed at the ER value. In Figure \ref{Fig:ER abund compared}b, we present the ACIS spectrum from the same epoch, with the Fe abundance associated with the hard model component allowed to vary.  Fitting the Fe abundance in the hard component model improves the broadband spectral fit ($\chi^{2}_{\nu} = 184/170$ to $\chi^{2}_{\nu} = 175/169$), for which an F-test indicates a marginal improvement (F-probablity $\sim$3 $\times$ 10$^{-3}$). While the overall broadband fit improvement by allowing the hard-component Fe abundance to vary is marginal, this may be misleading, as our feature of interest is a narrow-band region around Fe K emission line. We discuss the narrow-band at more detail in Section \ref{Sec:3.3}. It is rather clear that the improvement in the model fit is significant in the Fe K line band (Figure \ref{Fig:ER abund compared}b).

The measured Fe abundance associated with the hard component is a factor $\sim$2-3 higher than the previously measured Fe abundances of the ER with Chandra \citep{Zhekov09, Ravi21}. However, this abundance enhancement has large measurement uncertainties and could be consistent with the average Fe abundances of LMC stars \citep[e.g.,][]{Urbaneja17, Romaniello22}, making it statistically marginal.

The associated evolution of the hard-component electron temperature during this period is shown in Figure \ref{Fig:kT hot}. We observe a marginal linear increase (within uncertainties) in the hard component post-shock electron temperature. The more energetic electrons can excite the Fe XXV line complex significantly, resulting in an increase in its line intensity. Our observations strengthen the case for the presence of a He-like Fe XXV line complex. We note that, if the X-ray emitting plasma is in NEI, then the emission from He-like triplets is enhanced. This is consistent with our assumptions that the NEI effects are strong in the X-ray emitting plasma in SNR 1987A. As electron temperatures and abundances are varied simultaneously, there can be cross-correlation between their results and thus it is important to implement more complex and independent models to better constrain the potentially increasing Fe-abundance in SNR 1987A. 

\subsection{X-ray Light Curves} \label{sec:3.2}

\begin{figure*} 
     \centering
     \hspace{-0.2cm}
\includegraphics[scale=1.1,keepaspectratio]{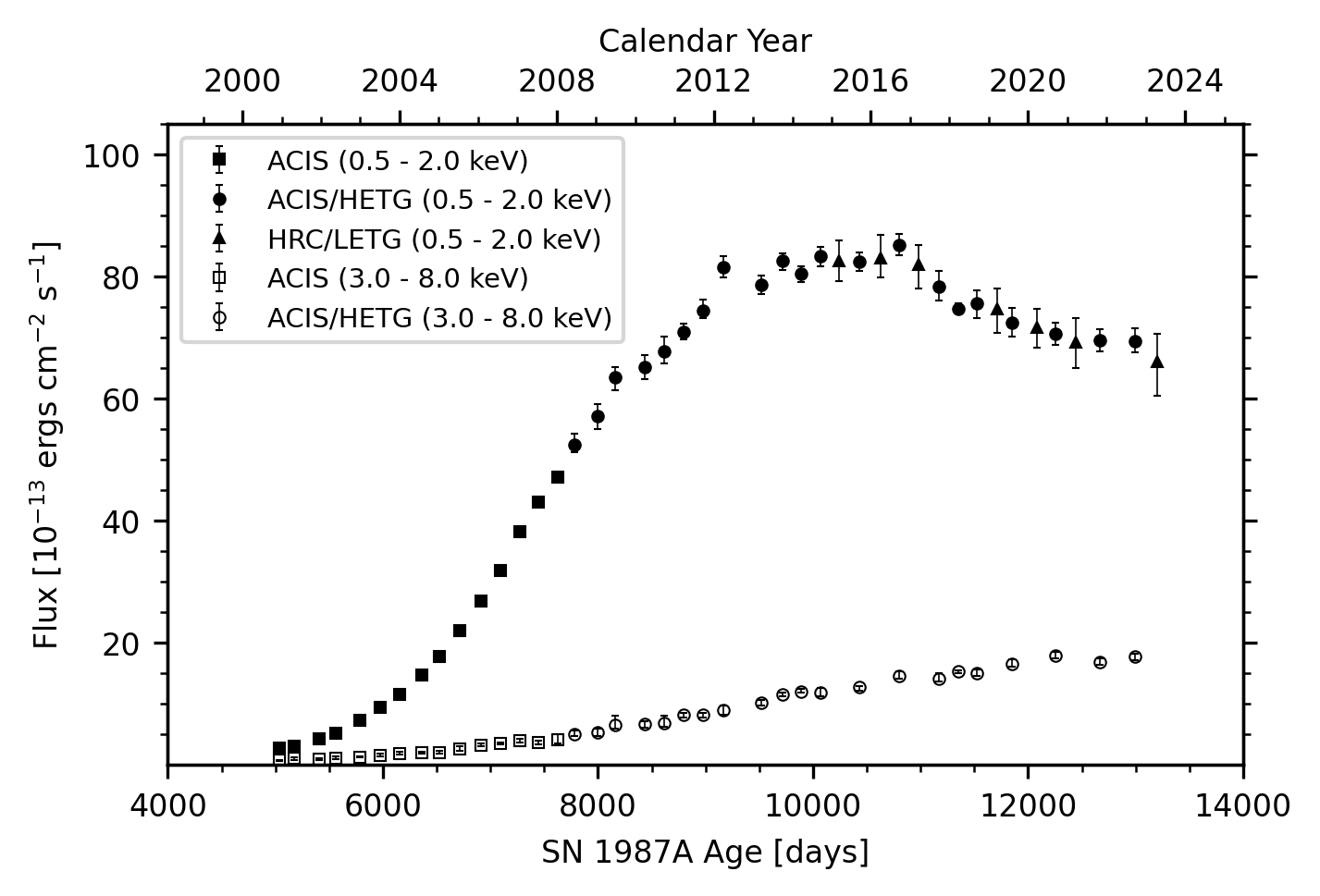} 

\caption{X-ray light curves of SNR 1987A between 1999 (5036 days) and 2023 (13,197 days). The soft-band (0.5--2.0 keV) fluxes are in filled markers and hard-band (3.0--8.0 keV) fluxes are in open markers. For several points, the error bars are too small to be visible.}
\label{Fig:X-ray LCs}
\end{figure*}

In Table \ref{Table:Observations}, we list our X-ray flux measurements based on our two-component spectral model fits to the ACIS and LETG spectra. An important systematic point of consideration in the ACIS flux measurements has been the temporal changes of the soft X-ray transmission of the ACIS OBF due to deposition of contaminants on its surface \citep{O'Dell13, Plucinsky18}. This contamination is strongest below 1 keV. To account for this artifact, we created the detector response maps for all of our new observations (taken since 2016) based on the calibration database with the latest ACIS OBF contamination models (see Section \ref{sec:2}).  We also re-created the detector responses for all previously-published data (taken before 2016) with the same calibration database for self-consistent flux measurements throughout the entire epochs.

To cross-check our soft X-ray flux measurements with the ACIS data, we measure the 0.5--2.0 keV band flux with the 1st-order HRC-S/LETG spectra, which is free of both pile-up and contamination. The measured fluxes from the LETG spectra between 2015 and 2023 are consistent with those that we measured based on our pileup-corrected ACIS spectra at corresponding epochs (within uncertainties), offering an independent verification and confidence in the true nature of the soft X-ray light curve. 
\setcounter{table}{1}
\begin{table*} [htbp!] 
\hspace{-0.2cm}
\caption{Fe K Line Profiles}
\begin{center}
\begin{tabular}{cccccc}
\hline \hline
Epoch & Age & Line Center & Line Equivalent Width & Line Flux   \\
 & (days) & (keV) & (keV) & (10$^{-6}$ photons cm$^{-2}$ s$^{-1}$) \\
\hline
Mar 2018 & 11351 & 6.65 $\pm$ 0.05 & 0.27 $\pm$ 0.08 & 3.9 $\pm$ 1.2\\
Sep 2018 & 11527 & 6.70 $\pm$ 0.06 & 0.52 $\pm$ 0.18 & 7.0 $\pm$ 2.9\\
Sep 2019 & 11849 & 6.72 $\pm$ 0.08 & 0.46 $\pm$ 0.16 & 7.3 $\pm$ 2.7\\
Sep 2020 & 12255 & 6.63 $\pm$ 0.09 & 0.41 $\pm$ 0.16 & 6.7 $\pm$ 2.6\\
Oct 2021 & 12666 & 6.69 $\pm$ 0.06 & 0.38 $\pm$ 0.15 & 6.3 $\pm$ 2.3\\
Sep 2022 & 12995 & 6.61 $\pm$ 0.05 & 0.50 $\pm$ 0.13 & 8.9 $\pm$ 2.7\\
\hline
\end{tabular}

\begin{tablenotes}
      \small
      \item All measured uncertainties are 90\,\% confidence intervals
\end{tablenotes}
\end{center}
\label{Table:Gaussian Fits Fe K}
\end{table*}

The resulting 0.5--2.0 keV (soft) and 3.0--8.0 keV (hard) X-ray light curves are presented in Figure \ref{Fig:X-ray LCs}. All measured X-ray fluxes between 1999 and 2016 are consistent with the previously published results in \cite{Frank16}, within statistical uncertainties. We find that the evolution of the soft X-ray light curve has changed significantly since 2016. The measured soft X-ray fluxes between 2016 ($\sim$10,800 days) and 2020 ($\sim$12,200 days) decline roughly linearly (by $\sim$4.5\,\% per year; Figure \ref{Fig:X-ray LCs}). Such a declining trend is generally consistent with the shock recently entering a low-density gas between 2011 and 2018, based on deep high-resolution Chandra HETG spectra \citep{Ravi21}. This recently declining soft X-ray flux has also been reported by the XMM observations during similar epochs \citep{Sun_2021}. Since 2020, the soft X-ray flux appears to level off at around $\sim$7 $\times$ 10$^{-12}$ erg s$^{-1}$ cm$^{-2}$ (Figure \ref{Fig:X-ray LCs}). Upcoming Chandra monitoring observations will be crucial to test this latest trend in the evolution of the soft X-ray light curve. 

In contrast, the hard X-ray light curve shows a steady increase until $\sim$2020 ($\sim$12,200 days). Since 2020, the observed hard X-ray fluxes might show hints of leveling off. However, it is difficult to draw a firm conclusion on this latest change in the hard X-ray light curve. Follow-up Chandra monitoring observations are required to test the true shape of the hard X-ray light curve of SNR 1987A.

\subsection{Fe K Emission Line} \label{Sec:3.3}
\begin{figure} 
     \hspace{-0.8cm}
\includegraphics[scale=0.355,keepaspectratio,angle=270]{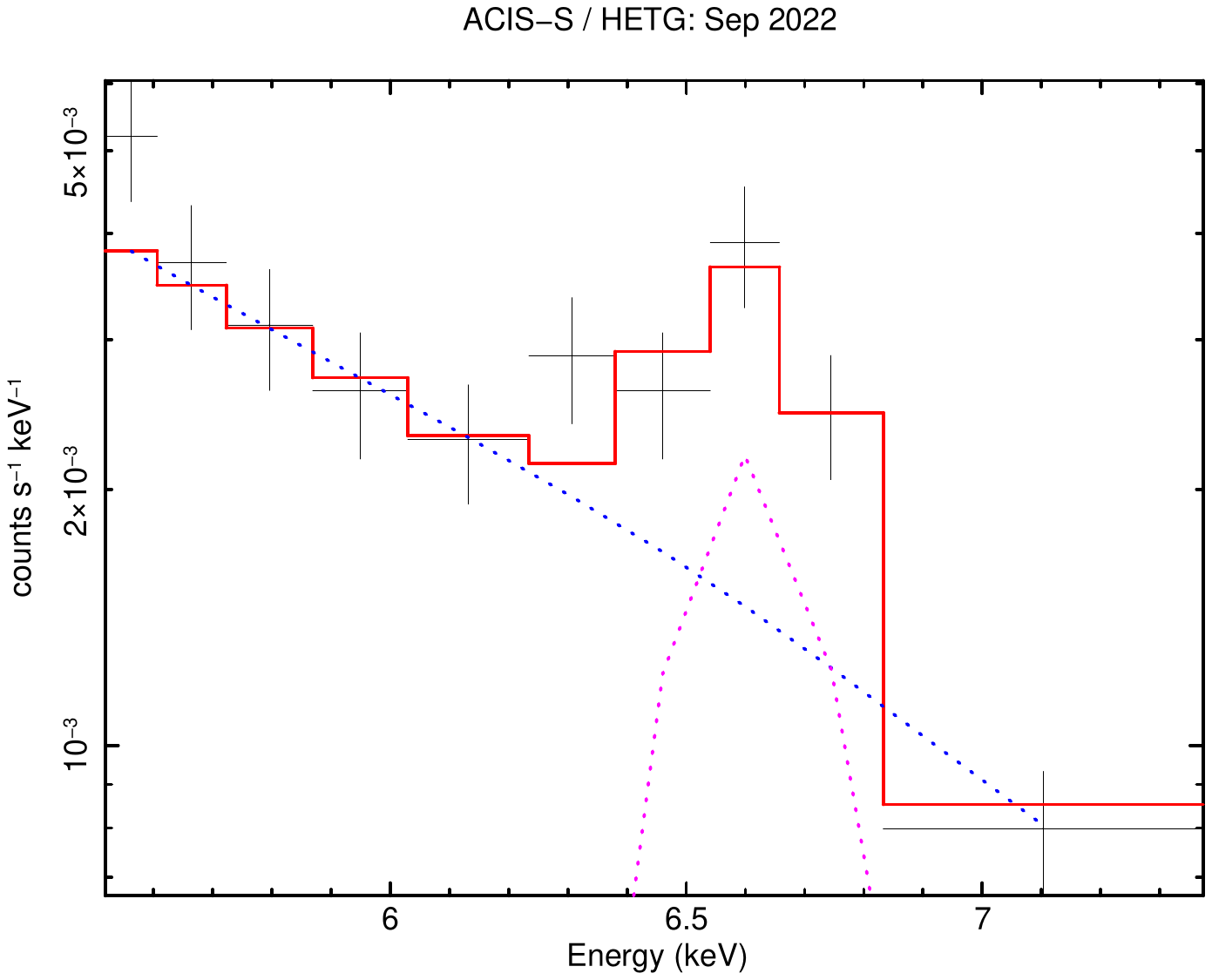} 
\caption{Absorbed powerlaw + Gaussian fits to the Fe K line observed in September 2022. The underlying continuum (characterized by a power law) component is in blue and the Gaussian component in magenta. The overall best-fit model is in red.}
\label{Fig:Sep_22_Fe_K}
\end{figure}

\begin{figure} 
\includegraphics[angle=90,scale=0.64,keepaspectratio,angle=270]{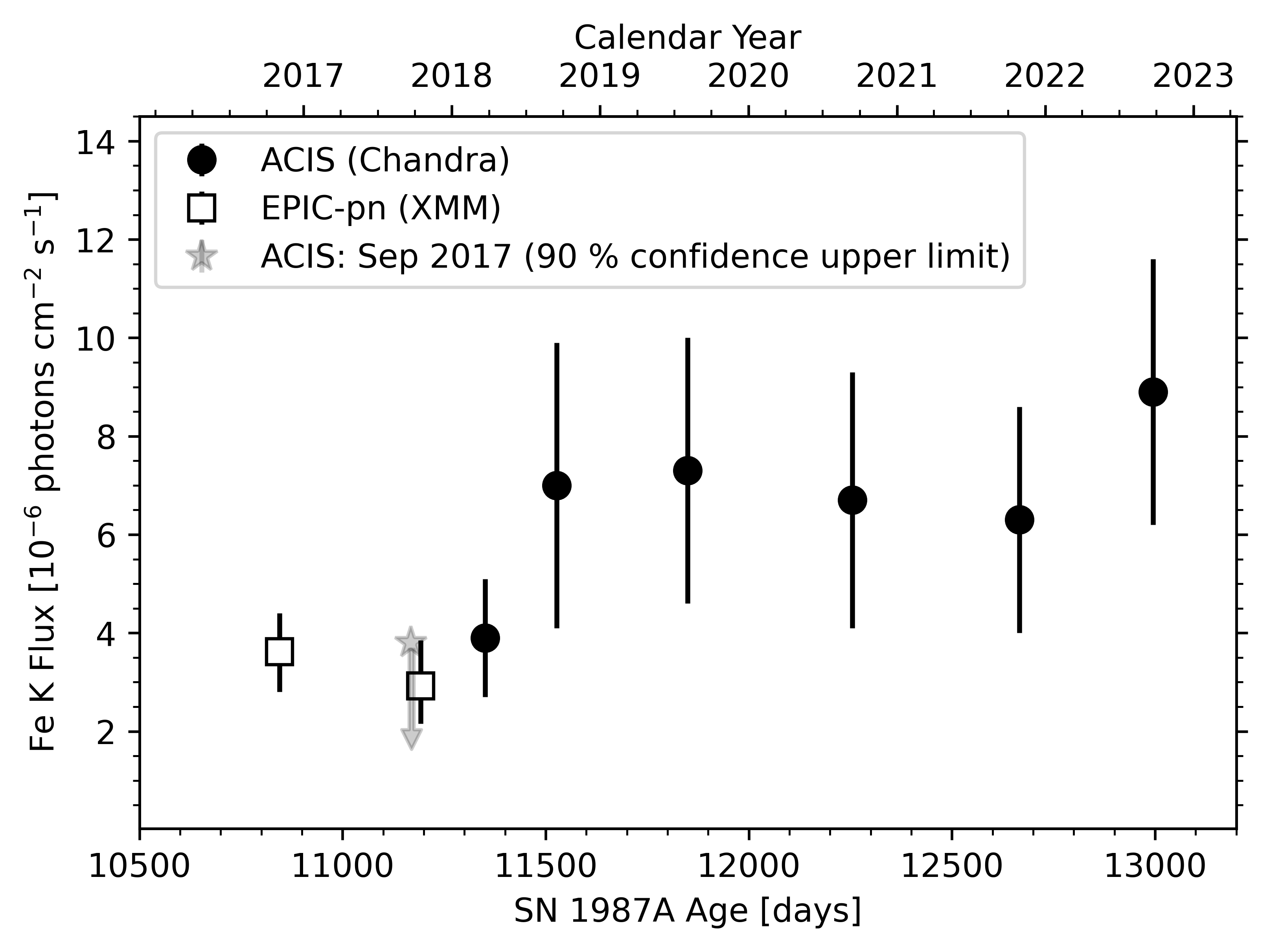} 
\caption{Measured Fe K line fluxes from ACIS 0th-order spectra between 2018 and 2022 (Table \ref{Table:Gaussian Fits Fe K}). Two epochs of EPIC-pn (XMM) measurements from \cite{Sun_2021} are also plotted. In September 2017 (day $\sim$11168), only an upper limit is achieved from the ACIS data.}
\label{Fig:Fe K flux}
\end{figure}
The detection of Fe K emission lines has been previously reported by several studies with XMM and NuSTAR \citep[e.g.,][]{Strum10, Maggi12, Sun_2021, Alp21}. These lines were not detectable in Chandra data prior to 2016 \citep{Frank16}. In this work, we report the detection of an excess above the continuum around $\sim$6.7 keV in the Chandra ACIS spectra, clearly identified first in March 2018 and then detected in every subsequent observation (Figure \ref{Fig:spectral fits}, \ref{Fig:Sep_22_Fe_K}).

To study the Fe K line profile and its temporal evolution, we fitted the 5.0--8.0 keV band ACIS spectra taken from March 2018 to September 2022 with an absorbed power law and a Gaussian model. Our best-fit model of the Fe K line profile in September 2022 is shown in Figure \ref{Fig:Sep_22_Fe_K}. Our best-fit line center, equivalent width, and line flux of the Fe K complex in each epoch between 2018 and 2022 are summarized in Table \ref{Table:Gaussian Fits Fe K}. The evolution of the Fe K line flux is presented in Figure \ref{Fig:Fe K flux}. Even though the measured uncertainties are large (typically $\sim$3$\sigma$ detections), the Fe K line detection has been significant ever since 2018. We also include a few representative epochs of Fe K line fluxes measured with EPIC-pn (on XMM-Newton) between 2015 and 2017 \citep{Sun_2021}. Our measured Fe K line flux in September 2017 shows large statistical uncertainties, and thus we place an upper limit (90\,\% confidence). We find a plausible agreement between the March 2018 Chandra and the October 2017 XMM result. Since 2018, our Chandra-measured Fe K fluxes significantly increase. Concurrent XMM fluxes are not currently available in the literature but future works will be useful to compare with our Chandra results.

We also measure the line equivalent width based on our best-fit Gaussian model-fits (Table \ref{Table:Gaussian Fits Fe K}). All of our measurements of the Fe K line fluxes, line equivalent widths, and the Fe abundances consistently indicate the recent development of strengthening X-ray emission from the shocked hot Fe gas. 
\begin{figure*} 
     \centering
     \hspace{-0.2cm}
\begin{longtable*}{c}
\includegraphics[width=\textwidth,keepaspectratio]{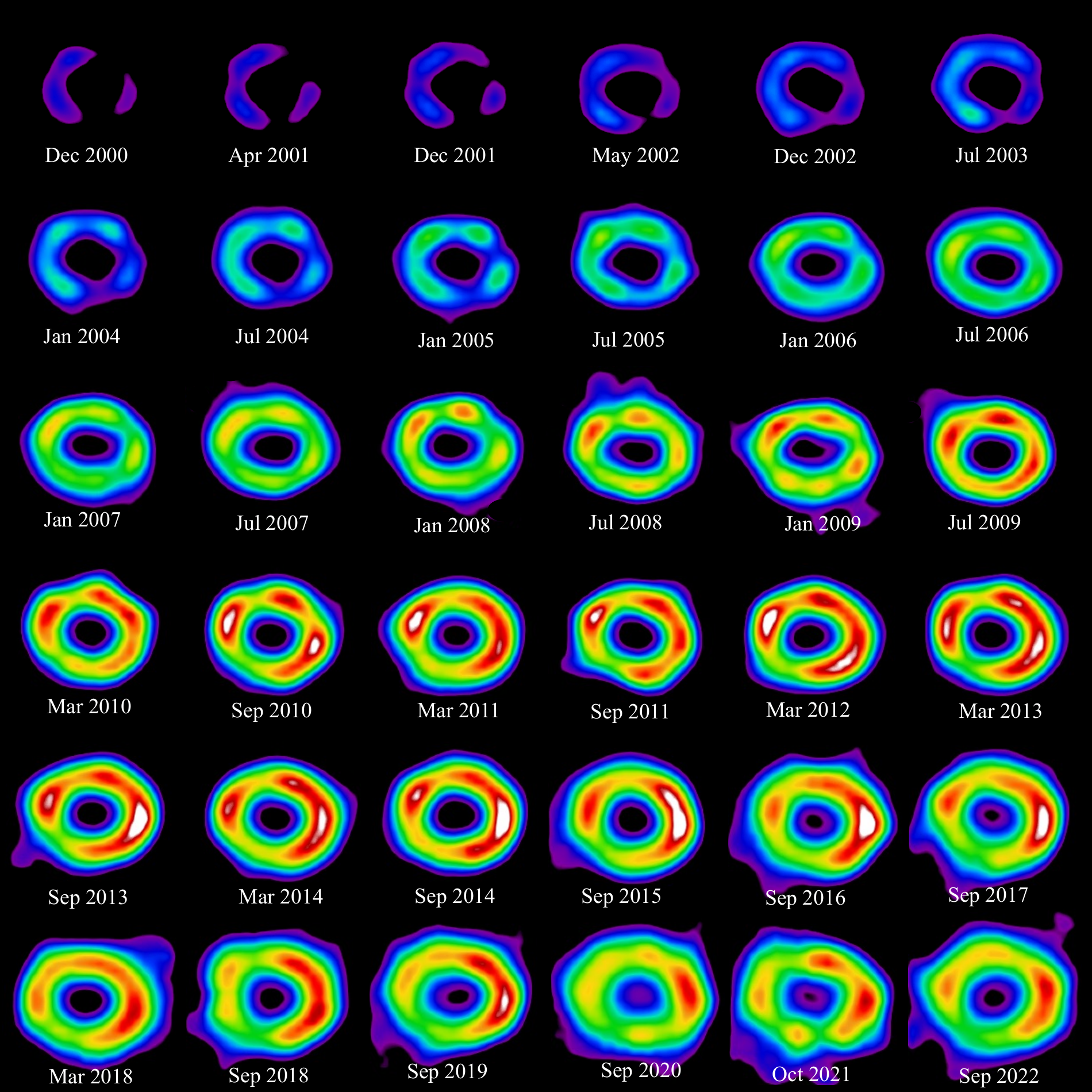} \\
\includegraphics[width=\textwidth,keepaspectratio]{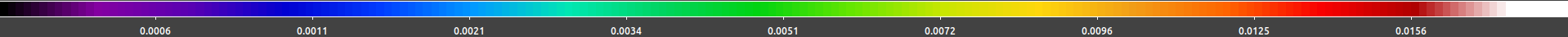}
\end{longtable*}
\caption{Deconvolved and smoothed broadband (0.3--8.0 keV) false-color Chandra ACIS images of SNR 1987A between December 2000 (day 5036) and September 2022 (day 12,995). All images are normalized by flux and use a square root scale on the same color scale (bottom). In all images, North is up and East is to the left.}

\label{Fig:images}
\end{figure*}

Measured mean line centroid energy of the Fe K profile in the ACIS spectra is 6.66 keV (Table \ref{Table:Gaussian Fits Fe K}). Based on XMM data, \cite{Sun_2021} estimated the Fe K line centroid to be 6.67 keV at 12,000 days, which is in good agreement with our Chandra measurements in 2019-2021 (Table \ref{Table:Gaussian Fits Fe K}). A line centroid energy of $\sim$6.7 keV approximately corresponds to the ionization state Fe XXV, consistent with a thermal origin \citep{Makishima86, Kallman04}. This identification of the Fe line is consistent with the evolution of electron temperatures associated with the hard component between 2016 and 2022 in our modeling (Figure \ref{Fig:kT hot}). It is also consistent with previously reported XMM-based measurements in \cite{Alp21}. In general, the overall observed Fe K complex is most likely a resultant from contributions of differently ionized Fe species. Thus, our results may indicate that the true evolution of Fe K line is due to a progressive ionization from the earlier detection by \cite{Maggi12}.

\subsection{X-ray Images} \label{sec:3.4}
\begin{figure} 
     \centering
     \hspace{-0.2cm}
\begin{longtable}{c}
\includegraphics[width=0.5\textwidth,keepaspectratio]{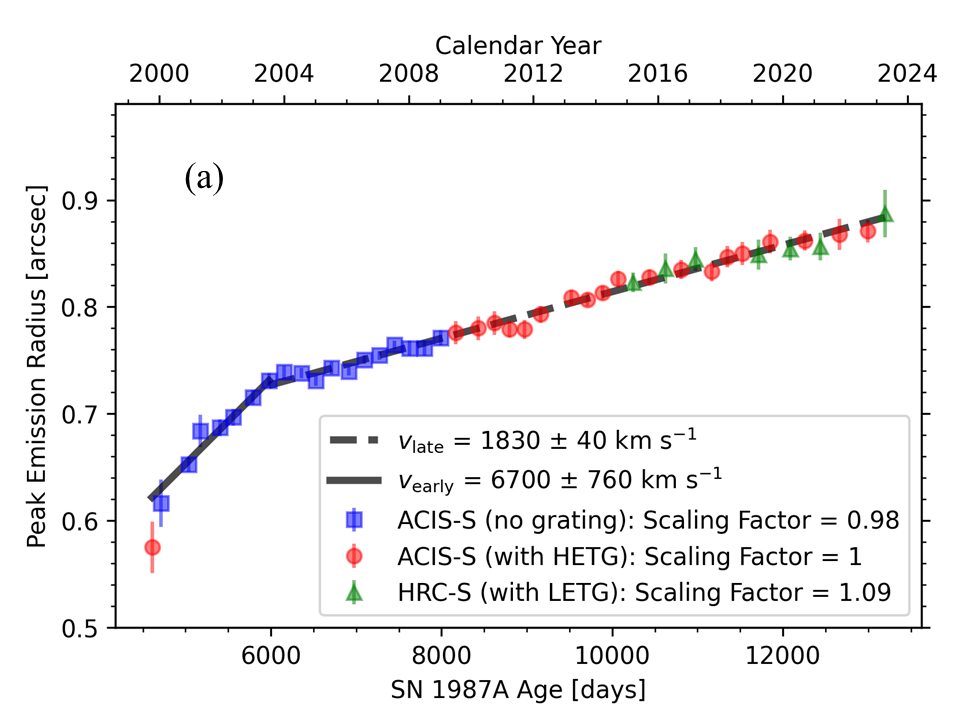} \\ \includegraphics[width=0.5\textwidth,keepaspectratio]{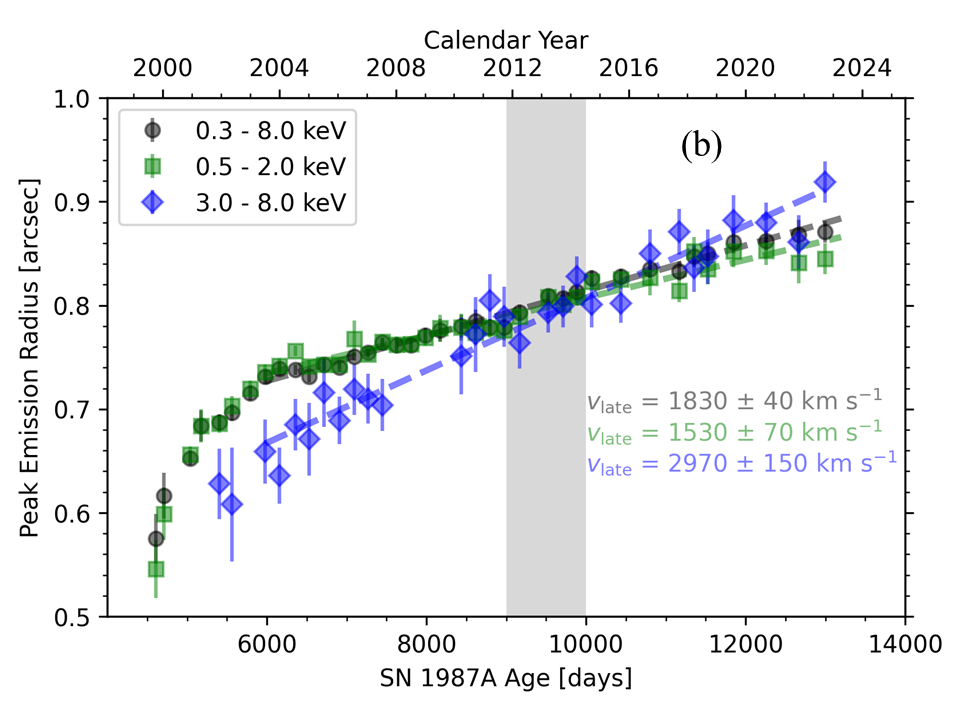} \\
\end{longtable}
\caption{(a): Best-fit radii of the 0.3--8.0 keV broadband ACIS images and HRC images between 1999 and 2023. A broken linear fit describes the expansion rate before and after the shock hitting the main body of the ER in 2004. Measured bare ACIS radii are scaled down by $\sim$2\,\% and HRC radii are scaled up by $\sim$9\,\%, which accounts for the cross-calibration in the radius measurements between bare ACIS and ACIS-S/HETG configurations and HRC-S/LETG and ACIS-S/HETG configurations, respectively (see text - Section \ref{sec:3.5}).  (b): Comparisons of the best-fit radii of SNR 1987A among the soft (0.5--2.0 keV), hard (3.0--8.0 keV), and the broadband (0.3--8.0 keV). A vertical band (grey) shows the period between $\sim$2012--2015, where the soft, hard, and broad energy band images have similar best-fit radii.}
\label{Fig:radial expansion}
\end{figure}

The superb angular resolution of Chandra offers a unique opportunity to resolve and study the changing morphology of SNR 1987A in X-rays. The evolution of Chandra ACIS images between 2000 and 2022 is shown in Figure \ref{Fig:images}. The ER was initially brighter in the east, but became brighter in the west after $\sim$2010 ($\sim$8500 days after SN) \citep{Park11, Frank16}. The change in the E-W brightness of the X-ray remnant is consistent with the temporal sequence of optical spots \citep[e.g.,][]{Sugerman02, Fransson15}, which supports the asymmetric propagation of the blast wave between east and west. 
The recent ACIS image in 2022 ($\sim$13,000 days) suggests that the western half has also dimmed since 2016, which is generally consistent with the declining soft X-ray light curve for the same period (Figure \ref{Fig:X-ray LCs}). 
\begin{figure} 
     \centering
     \hspace{-0.2cm}
\begin{longtable}{c}
\includegraphics[width=0.5\textwidth,keepaspectratio]{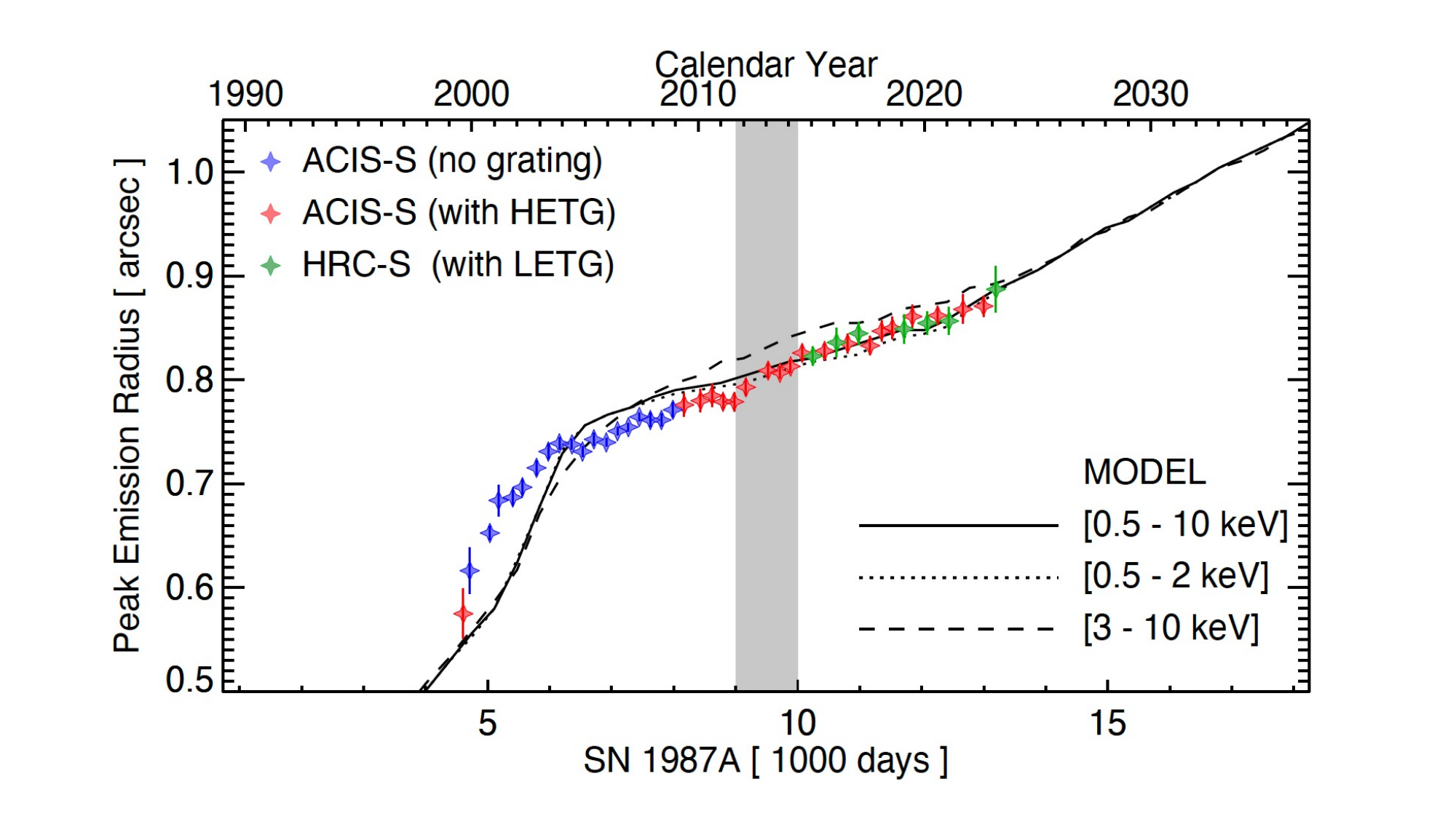} 
\end{longtable}
\caption{Comparison between best-fit radii of the 0.3--8.0 keV broadband ACIS images and HRC images between 1999 and 2023 with MHD model-predicted expansion rates in 0.5--10.0 keV (solid), 0.5--2.0 keV (dotted), and 3.0--10.0 keV (dashed) energy bands. A vertical band (grey) shows the period between $\sim$2012--2015, where the soft, hard, and broad energy band observed images have similar radii. General trends among the model expansion rates are consistent with observations.}
\label{Fig:radial expansion model}
\end{figure}

\subsection{Radial Expansion Rate} \label{sec:3.5}
The high cadence of ACIS observations, coupled with excellent spatial resolution, enables us to track the changes in the X-ray radial expansion of SNR 1987A. We measure the radius of the X-ray emitting ring and its thickness by adopting similar methods to those described in \cite{Racusin09}. In these methods, we fit each deconvolved ACIS image to an image model that consists of four representative lobes and an underlying uniform ring. The estimated radius represents the average distance from the center to the peak X-ray emission in SNR 1987A. The thickness of the independent lobes is a free parameter in our image model, which is averaged out to get the width of the X-ray emitting ring.

The best-fit radii of the broadband images are shown in Figure \ref{Fig:radial expansion}a and Table \ref{Table:Observations}. We also measured the radius for soft- and hard-band images to study the energy dependence of the X-ray radius (Figure \ref{Fig:radial expansion}b). For each band, epochs with at least several hundred counts were included in the radial expansion analysis to ensure statistically-reliable measurements. Applying a similar image modeling to that we used for the ACIS images \citep[e.g.,][]{Racusin09, Park11, Helder13, Frank16}, we measured the SNR's radii based on our 0th-order HRC data for several epochs between 2015 and 2023 (Table \ref{Table:Observations}). 

Because three different detector modes (bare-ACIS, ACIS/HETG 0th-order, and HRC/LETG 0th order) are used to measure the angular size of SNR 1987A down to $\sim$arcsec scales, we cross-calibrate to test any systematic effects in our measurements among these configurations. Measured radii with bare ACIS configuration are systematically higher (by $\sim$2\,\%) than the corresponding ACIS-S/HETG configurations. This small systematic effect was not identified in our previous work \citep{Frank16}. While \cite{Frank16} used an older version of PSF created by MARX, we used the up-to-date PSF created by ChaRT \citep[using SAOTrace\footnote{https://cxc.harvard.edu/cal/Hrma/SAOTrace.html}][]{Carter03}, which is probably a more accurate description. SAOTrace is a high-fidelity ray-tracer for the Chandra mirrors that simulates many details that are treated statistically in MARX to improve efficiency. The two Chandra epochs between 2008 and 2009 (when HETG was first inserted) were observed with both the bare ACIS and the ACIS-S/HETG configurations (ObsIDs: 9143, 9144 and ObsIDs: 10130, 10855 - Table \ref{Table:Observations}). We estimated the cross-calibration factor between ACIS and ACIS-S/HETG configurations based on radii measurements at these epochs.

Measured radii with HRC are systematically lower than their corresponding ACIS epochs (by $\sim$9\,\%), indicating systematic effects between two different types of detectors (including differences in their spatial resolutions). Nonetheless, the SNR's expansion rate in HRC images is consistent with that in ACIS images (within statistical uncertainties) between 2016 and 2022, providing an independent confirmation of the radial expansion rate of SNR 1987A that we measure with the ACIS images (see below). The HRC images are scaled up by $\sim$9\,\% and fit together with the ACIS images to estimate the late-time (since $\sim$2004) velocity, $v_{late}$ in Figure \ref{Fig:radial expansion}a.

We estimate the early velocity, $v_{early}$ of the expanding broadband images before the strong shock interaction with the ER at around 2004 as 6700 $\pm$ 750 km s$^{-1}$, and it reduces to 1830 $\pm$ 40 km s$^{-1}$ after 2004. Our refined measurements of expansion rates are consistent (within uncertainties) with earlier results of \cite{Racusin09}, \cite{Helder13}, and \cite{Frank16}. Since 2016, the overall expansion rate of SNR 1987A continues to show a linear rate similar to that estimated for the period of 2004-2015. Thus, the bulk of broadband X-ray emission still appears to be dominated by emission from the shocked dense CSM in the ER until 2022. 

In Figure \ref{Fig:radial expansion}b we compare the energy dependence of the expansion rate between soft, hard, and the broadband ACIS images.  The soft band expansion (in green) is similar to the broadband expansion (in black) until $\sim$2016, indicating that the soft band emission was dominating the broadband image. Since $\sim$2016, the soft band radius appears to become smaller than the corresponding broadband radius. In fact, our linear expansion model fits show that the expansion rate (1530 $\pm$ 70 km s$^{-1}$) in the soft band images is slightly lower than that (1830 $\pm$ 40 km s$^{-1}$) in the broadband images (Figure \ref{Fig:radial expansion}b). 

The hardband expansion rate has been steeper (than soft/broad band rates) throughout the entire period of our measurements. The radius of the hard band image was significantly smaller compared to respective broadband images at early epochs, but it caught up with the broad/soft band size in $\sim$2012--2015 ($\sim$9000--10,000 days, Figure \ref{Fig:radial expansion}b). This epoch is coincident with the period when the soft X-ray light curve leveled off (Figure \ref{Fig:X-ray LCs}). By 2018 ($\sim$10,300 days), the hard band images have started to overtake the broadband emission size. In 2022 ($\sim$13,000 days), the hard band image is clearly larger than either of the soft or broad X-ray images of SNR 1987A (Figure \ref{Fig:radial expansion}b). A fit between 2004 and 2022 of the hard band images gives an expansion rate of 2970 $\pm$ 150 km s$^{-1}$, which is significantly faster than the peak emission expansion rate observed from the broadband images for the same time period. 

In Figure \ref{Fig:radial expansion model}, we compare the radial expansion rates from observations (ACIS broadband images and HRC images) with the MHD model-predicted expansion rates. The radius in the model hard band (3--10 keV) is initially smaller than in the model soft band (0.5--2.0 keV), eventually becoming larger like their observed counterparts. The slope in the model hard band during the interaction with the dense ring is steeper than in the model soft band because the shock velocity in the H II region and less dense inter-clump material (which mainly contributes to the hard X-ray emission) is higher than the shock velocity in the densest component of the ring (contributing to the soft X-ray emission). The soft model component closely tracks the broadband model component, suggesting that peak X-ray emission is still dominated by the dense ER. Thus, predicted MHD model evolution of the X-ray radii at different energies is consistent with the observed sub-band trends (Figure \ref{Fig:radial expansion}). The discrepancies present between model-predicted radii and observations could be due to the assumed density structure of CSM in the model, and our observations will be useful in constraining these assumptions. However, qualitatively, the model replicates our observed trends, supporting the physical picture of SNR 1987A's evolution. The model further predicts a notably steeper expansion rate both presently and in the future. This is attributed to the faster expansion of the remnant in the less dense CSM beyond the ring. Once again, this aligns with the physical picture emerging from our observations.

\section{Discussion \label{sec:4}} 

\subsection{X-ray Emitting Plasma: Light Curves and Fe Abundance} \label{sec:4.1}
The physical conditions of the X-ray emitting plasma have been approximately modeled with a characteristic bimodal temperature distribution of plane parallel shocks in the literature \citep[e.g.,][]{Frank16}. The cool ($kT$ = 0.3--0.9 keV) or soft component represents X-ray emission primarily from the shocked dense clumpy materials in the ER, while the hot ($kT$ = 1.5--3.0 keV) or hard component describes the combination of X-ray emission from within the inter-clump gas in the ER and from the shocked less-dense material from the H II region \citep{Zhekov10, Orlando15}.

\begin{figure} 
     \hspace{-0.2cm}
\includegraphics[width=0.5\textwidth,keepaspectratio]{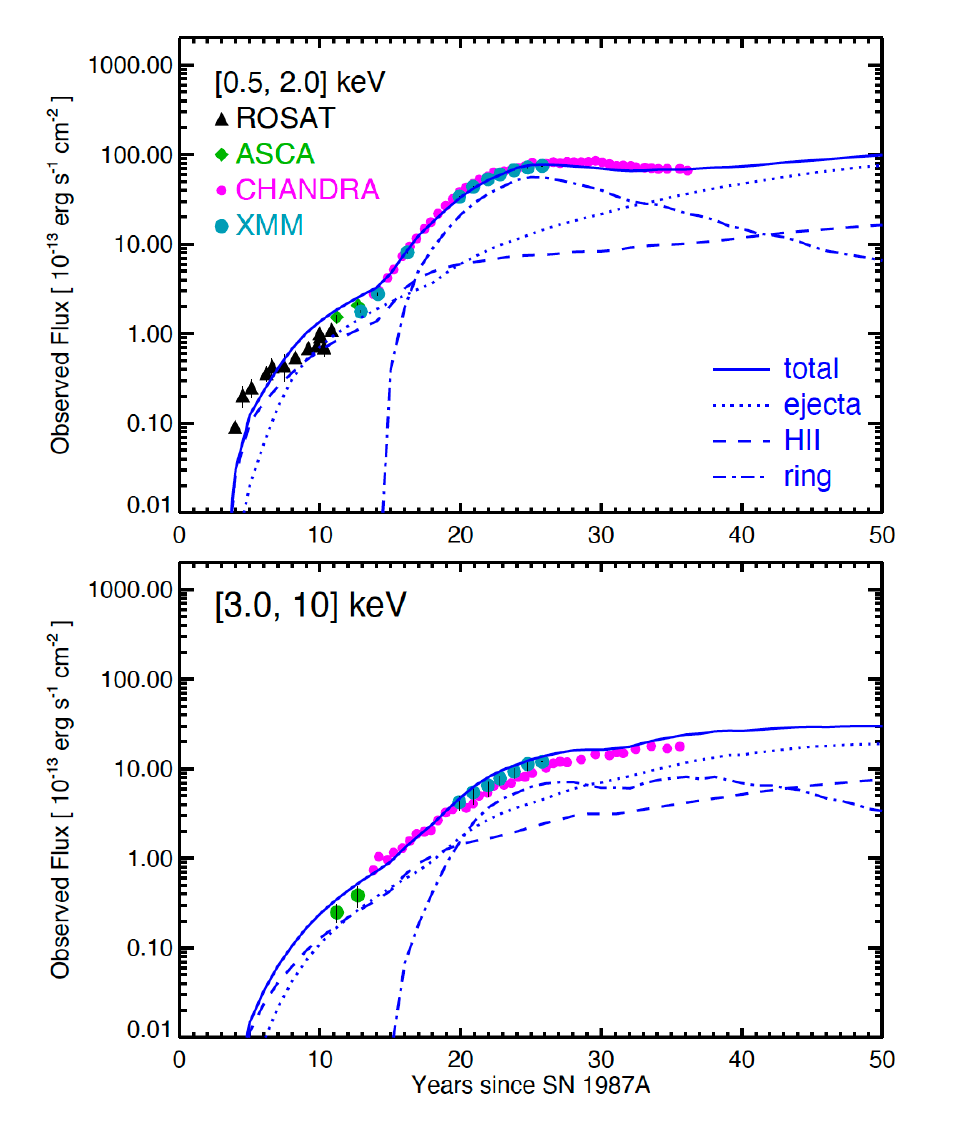} 
\caption{\textbf{Top}: Comparison between observed soft X-ray fluxes and corresponding MHD-simulated model components \citep{Orlando20}. Chandra fluxes (magenta) are from this work, XMM fluxes (blue) from \cite{Haberl06, Maggi12}, ROSAT fluxes (gray) from \cite{Haberl06}, and ASCA fluxes (brown) from Appendix B of \cite{Orlando15}. The MHD model predicts that the soft X-ray flux from the shocked ejecta (green) may become more significant than that from the shocked ER (red) in $\gtrsim$2022 ($\gtrsim$35 yr). \textbf{Bottom}: Similar comparison between observed hard X-ray fluxes and the corresponding MHD-simulated model components. }
\label{Fig:mhd_lc}
\end{figure}

Since 2011, the soft X-ray flux from SNR 1987A had levelled off at $\sim$8 $\times$ 10$^{-12}$ erg s$^{-1}$ cm$^{-2}$ \citep{Frank16}. Such a flattening has been previously predicted to be an indicator of the blast wave starting to move out of the dense ER \citep{Park11, Dewey12, Orlando15}. We find that the soft X-ray light curve started to decline roughly linearly since 2016, decreasing to $\sim$7 $\times$ 10$^{-12}$ erg s$^{-1}$ cm$^{-2}$ by $\sim$2020. This new feature in the soft X-ray light curve is consistent with the physical interpretation in which the forward shock's propagation beyond the ER has become significant, heating hitherto unknown low-density CSM. Since 2020, we find that the soft X-ray light curve appears to have stopped declining, and leveled off. This latest development in the soft X-ray light curve of SNR 1987A is in plausible agreement with the MHD model predictions \citep[][and Figure \ref{Fig:mhd_lc}]{Orlando20}, in which the emerging X-ray emission from the reverse-shocked ejecta is expected to cause a similar change in the soft X-ray light curve. 

The MHD model predicted that the X-ray flux from the reverse-shocked outer layers of ejecta would grow to be the dominant component in the soft X-ray spectrum in $\sim$2022 or later. Based on these MHD models, the reverse shock is currently traveling through the extended H-envelope and the underlying He-envelope. According to the stellar evolution model used in these SN-SNR simulations, the mass of these H and He envelopes is about 10 M$_{\odot}$ \citep{Ono20}. So, the shocked ejecta in the MHD models are mainly shocked material from these envelopes. As the reverse shock travels further into the ejecta, significant changes in the X-ray spectrum of SNR 1987A (i.e., developments of strong emission lines from the shocked heavy ions such as Si, Fe, O etc.) are expected in near future due to the expected dominating contribution from the reverse-shocked metal-rich ejecta. 

Our observations of the increasing Fe K line flux may indicate the onset of X-ray emission from the reverse-shocked ejecta. However, the current results from Chandra data are insufficient to distinguish them from a CSM origin. Comparing with MHD simulations \citep{Orlando20}, we find that the contributions from both the ejecta component and the ring component (CSM) are comparable at this stage (see Figure \ref{Fig:mhd_lc}). These simulations suggest that the shocked ejecta originate from the outer layers rich in H and He, while the Fe-rich ejecta in the innermost layers exhibit significant asymmetry due to the bipolar explosion \citep{Ono20, Orlando20}. This asymmetry results in the extension of Fe-rich plumes through the outer layers of ejecta, bringing them closer to the reverse shock. Current models predict that these Fe-rich plumes have not yet started the interaction with the reverse shock. However, recent observations with JWST NIRSpec of SNR 1987A have unequivocally detected this asymmetry\footnote{Previously, \cite{Larsson16} discussed the ejecta asymmetry of SNR 1987A at optical and near-infrared wavelengths.} and suggested that the Fe-rich plumes have already begun interacting with the reverse shock \citep{Larsson23}. Follow-up Chandra monitoring observations are therefore essential to validate these observed trends at X-ray wavelengths and provide observational constraints to improve the explosion models.

\begin{figure} 
     \hspace{-0.2cm}
\includegraphics[width=0.5\textwidth,keepaspectratio]{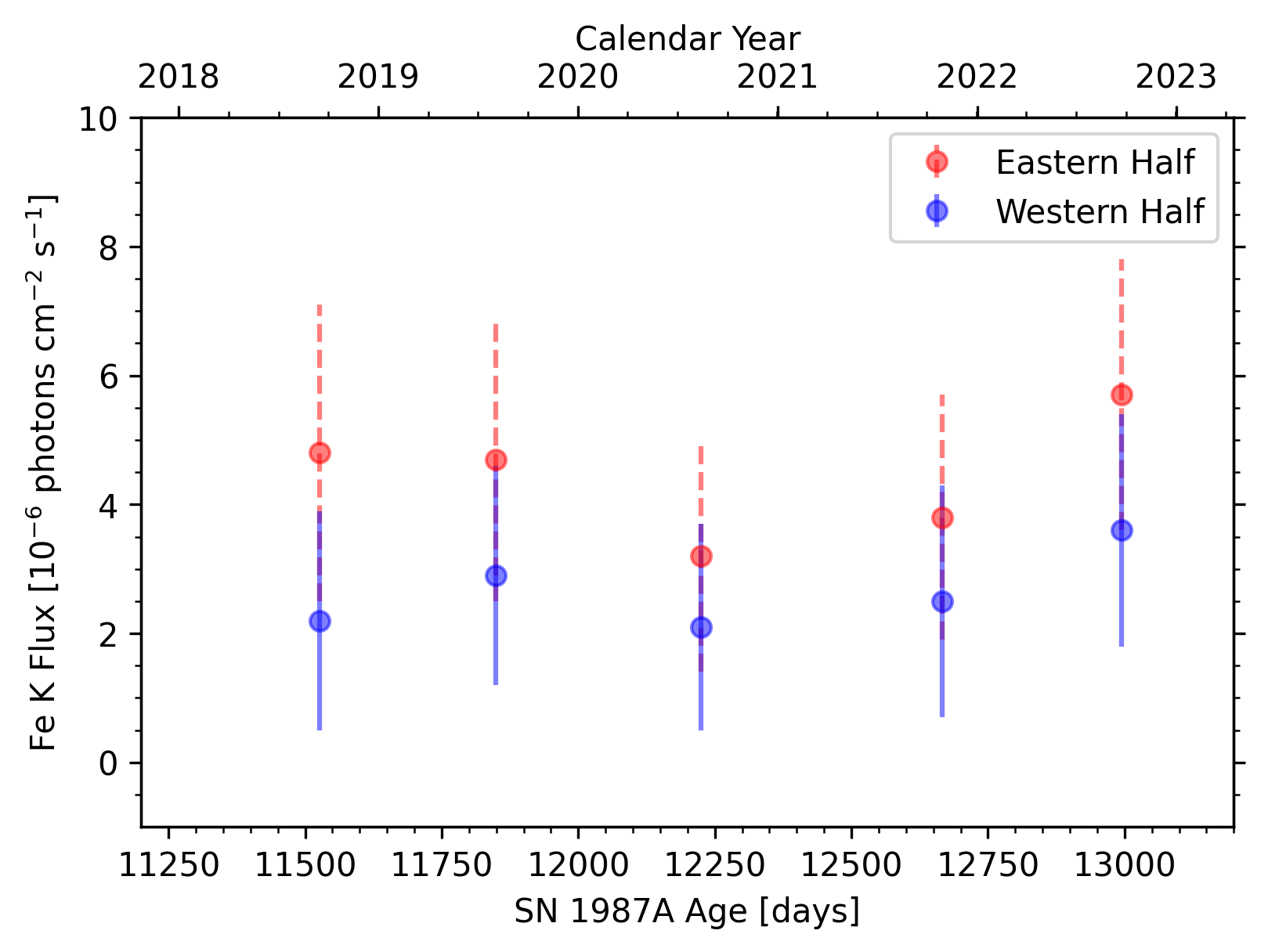} 
\caption{Fe K flux measured from the E and W-halves of SNR 1987A between 2018 and 2022.}
\label{Fig:E-W asymmetry}
\end{figure}

\subsection{Asymmetry of Fe K Emission} \label{sec:4.2}
\begin{figure} 
     \hspace{-0.2cm}
\includegraphics[width=0.5\textwidth,keepaspectratio]{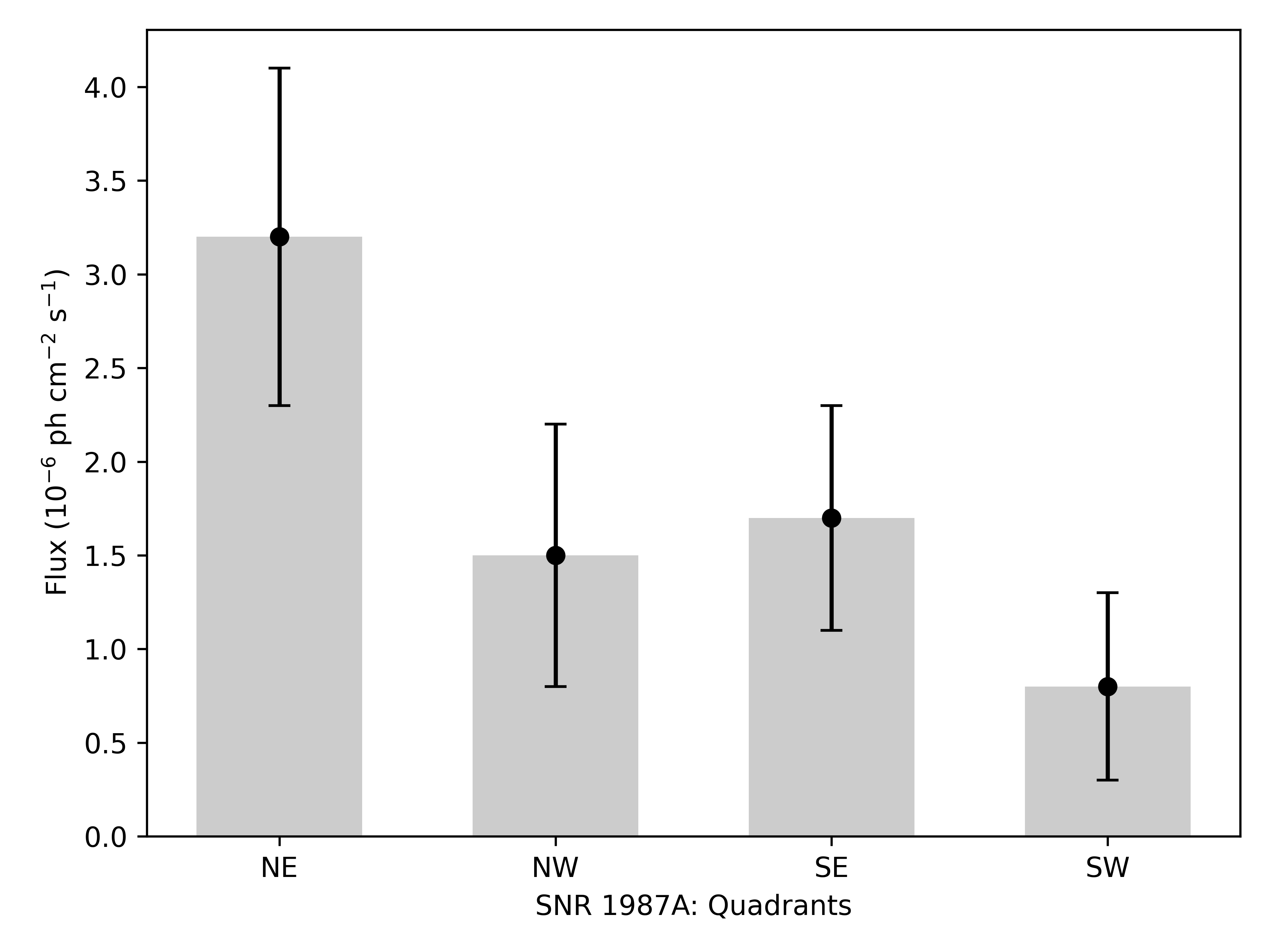}
\caption{Fe K flux measured in the combined spectra (between 2020 and 2022) from northeast (NE), northwest (NW), southeast (SE), and southwest (SW) quadrant regions of SNR 1987A. Fluxes are measured as described in Section \ref{sec:3.1}. Uncertainties are 90\,\% confidence intervals.}  
\label{Fig:Quadrant asymmetry}
\end{figure}

Chandra X-ray images showed the brightening in the eastern side first, and then in the western side subsequently (around 2010; Figure \ref{Fig:images}). Similar time sequences have been recognized in the developments of optical hot spots and radio lobes \citep{Sugerman02, Ng13, Zanardo14, Fransson15}. These multi-wavelength observations may indicate an asymmetric progress of the blast wave between eastern and western sides of the SN, probably due to an asymmetric explosion and/or non-uniform CSM distribution. In either scenario, an asymmetric progress of the reverse shocks between the eastern and western sides of the SNR would be expected, which may accordingly result in time-differential developments of the X-ray emission from the reverse-shocked ejecta between eastern and western halves of the SNR. 

Motivated by the recent development of the Fe K emission line, the hints of enhanced Fe abundance, and the JWST observations \citep{Larsson23}, we may entertain the scenario that the strengthening Fe K emission line is due to reverse-shocked ejecta. Considering the asymmetric developments of X-ray emission between the eastern and western regions of SNR 1987A, we may also expect an asymmetry in the Fe K line flux due to a similar asymmetric development of the reverse shock between E and W halves. 

To test this scenario, we measured the Fe K line fluxes from the eastern and western halves of SNR 1987A between 2018 and 2022 (Figure \ref{Fig:E-W asymmetry}). While uncertainties are large due to low count statistics in the Fe K line features extracted from eastern and western halves, it is interesting to find that best-fit Fe K line flux is $\sim$twice higher in the eastern half at all epochs.  This systematic spatial distribution of the Fe K line flux is supportive of our assumed physical picture of the Fe K line originating from the asymmetric reverse-shocked ejecta. However, the alternate scenario of enhanced Fe fluxes due to changing thermal conditions of an asymmetric CSM origin between E- and W-halves is equally feasible. As the contribution of the emerging emission from the shocked ejecta in the X-ray spectrum is expected to become more significant in the future, continuing Chandra observations will present a unique opportunity to test the hypothesis of such an asymmetric development of the reverse shock in SNR 1987A. 

We combine all ACIS data taken from 2020 to 2022 (where we see the enhanced Fe K line fluxes, Figures \ref{Fig:X-ray LCs} and \ref{Fig:mhd_lc})) to increase photon count statistics in the Fe K line feature to investigate a more detailed Fe K line flux asymmetry. We compare the Fe K emission feature in the combined spectrum from equally sized (radius $\sim$4 arcsec) northeast (NE), northwest (NW), southeast (SE), and southwest (SW) quadrant regions. We follow the same prescriptions described in Section \ref{sec:3.1} to perform a broadband spectral fit to measure the Fe-abundance associated with the hard component, and a narrow-band fit around the Fe K line (5.0--8.0 keV) to measure the emission line flux across the four quadrants.

In Figure \ref{Fig:Quadrant asymmetry}, we present the flux  distributions associated with Fe K emission across the four quadrants. We find that the strongest Fe K flux is observed in the NE quadrant. We also confirm that on average, the total flux from the eastern half is stronger than the western half, as expected from analyzing the integrated spectrum from each epoch (Figure \ref{Fig:E-W asymmetry}).  The significant flux difference between the NE and SW quadrant could suggest that X-ray emission from the reverse shock is first emerging in the NE quadrant. 

Recent NIRSpec JWST observations of SNR 1987A have revealed two primary Fe-rich plumes of ejecta, moving in the NE and SW directions \citep{Larsson23}. Such a blue-shifted NE plume and a red-shifted SW plume geometry of the Fe-ejecta was predicted by the MHD model described in \cite{Orlando20}. If the remnant interacts first with the ER in the eastern hemisphere, a reflected shock results earlier in the east than in the west, pushing the reverse shock in the eastern half to propagate through the innermost ejecta earlier than in the western side. This could potentially lead to the interaction of the reverse shock with the plume in the NE quadrant first, followed by the SW quadrant at a later time. The JWST NIRSpec images illustrate that the plume propagating to the NE lies in the plane of the ring, whereas the plume propagating to the SW does not \citep[see Figure 4 in][]{Larsson23}. Consequently, the majority of the material in the SW Fe-rich plume is situated farther away from the reverse shock. So, the possibility that the reverse shock is more internal in the eastern half than in the western half due to an earlier interaction with the ring, coupled with the fact that the NE plume is closer to the reverse shock (being planar with the ER) than the SW plume, makes it plausible that the interaction between Fe-rich ejecta and the reverse shock (radiating in X-rays) occurs first within the NE plume, consistent with our results (Figure \ref{Fig:Quadrant asymmetry}).

Future Chandra observations of the strengthening Fe K line would be useful for constraining the potential blue- and red-shifts among the observed Fe K line profiles in the NE and SW quadrants, respectively. If the Fe K feature is indeed strengthening with time after 2022, we expect better count statistics between $\sim$6--7 keV to be able to map out the spatial distribution of the Fe K line flux in comparison to that of the underlying continuum, for which Chandra data are uniquely suited. Testing for spatially-differential developments of the reverse shock between eastern and western sides of the SN and the robust measurements of enhanced Fe abundance with future high-resolution X-ray spectroscopy would also provide useful observational limits to constrain modeling of the explosion physics and the nature of the progenitor of SN 1987A.

\begin{figure} 
     \hspace{-0.2cm}
\includegraphics[width=0.5\textwidth,keepaspectratio]{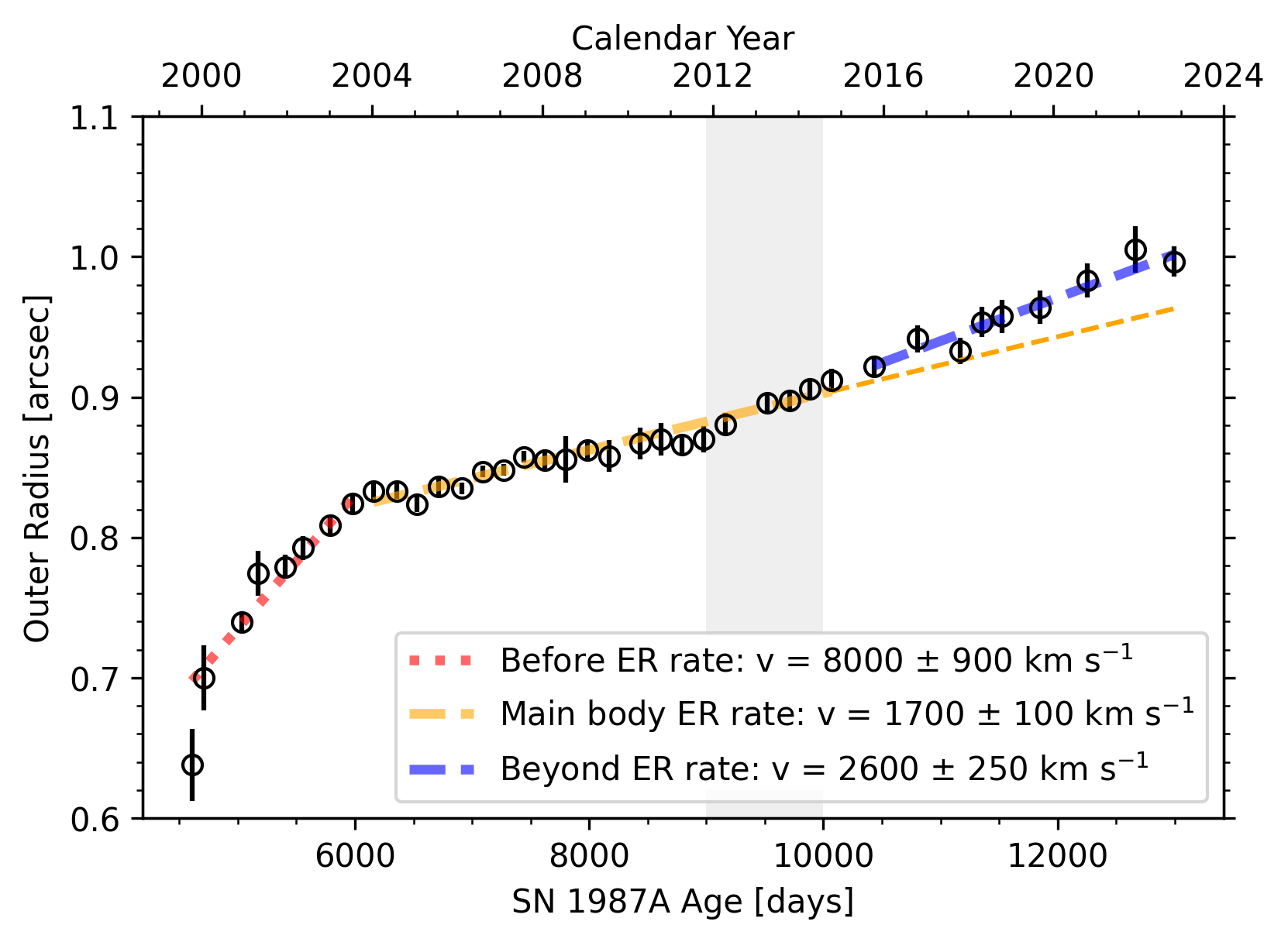} 
\caption{Evolution of the outer radius, $r_{outer} = r_{0} + \sigma_{0}/2$, where $r_{0}$ is the peak emission radius, and $\sigma_{0}$ is the best-fit width of the ring. We approximate the outer radius of SNR 1987A as the location of the blast wave. The best-fit linear expansion rates are overlaid for three epochs: before the shock enters the main body of the dense ER (red), during the shock interaction with the main body of the ER (yellow), and after the shock enters the low-density CSM beyond the main body of the ER (blue). The expansion rate during the shock interaction with the dense ER (orange dashed line) is extrapolated to compare with the expansion rate beyond the ER. The vertical shaded band (grey) between $\sim$2012--2016 marks the epochs when the size of the hard band X-ray SNR reached that of the soft band and broadband SNR (as shown in Figure \ref{Fig:radial expansion}).}
\label{Fig:outer_radius}
\end{figure}

\subsection{Evolution of the Forward Shock} \label{sec:4.3}

The overall radial expansion rate of SNR 1987A in the broadband images between 2016 and 2023 is approximately linear, similar to that between 2004 and 2016. This suggests that the peak X-ray emission from SNR 1987A is still dominated by the bright ER emission as of 2023. On the other hand, the width of the X-ray-emitting ring has increased to $\sim$0\farcs 27 in 2022, whereas it stayed at $\sim$0\farcs 20 until 2012. The width of the X-ray ring increased linearly from 2012 to 2022 at a rate of $\sim$4\,\% per year. While the net increase of $\sim$0\farcs 07 in the X-ray ring width from 2012 to 2022 is small and comparable with the ACIS astrometry limit, we note that our best-fit measurements of the ring width consistently show a systematic increase over a dozen epochs covering several years, while they were staying constant for the previous decade. Also the measured width in 2022 ($\sim$0 \farcs 27) is comparable to the effective spatial resolution of our ACIS images (see Section \ref{sec:3.2}). 
Thus, we conclude that the recent increase in the ring width over the past decade is real. Based on these results, we trace the radius of the outer boundary of the X-ray remnant with $r_{0} + \sigma_{0}/2$, where $r_{0}$ is the X-ray radius associated with the peak emission and  $\sigma_{0}$ is the width of the X-ray emitting ring in our image model fits (see Section \ref{sec:3.5}).

In Figure \ref{Fig:outer_radius}, we plot the outer radius of SNR 1987A. The outer radius was expanding at a speed of 8000 $\pm$ 900 km/s until 2004 when the forward shock entered the main body of the ER. This high expansion rate is similar (within uncertainties) to that measured for the peak X-ray emission \citep{Frank16}. It may indicate that before the forward shock hit the main body of the ER, the peak intensity and the outer boundary of the X-ray SNR 1987A were approximately co-moving at similar velocities. Once the shock entered the main body of the ER, the expansion rate of the outer boundary of the SNR slowed down to 1700 $\pm$ 100 km s$^{-1}$ while the shock interacted with the dense ER. This velocity is similar (within uncertainties) to the overall velocity of the peak emission during this period \citep[Figure \ref{Fig:radial expansion}a,][]{Frank16}, which may indicate a strong de-acceleration of the shock front (outer boundary) as it entered the main body of the ER.

Sometime between 2012 and 2016, the outer edge started to expand faster (2600 $\pm$ 250 km s$^{-1}$ - Figure \ref{Fig:outer_radius}), indicating that the shock was starting to move out of the ER. This epoch also coincides with the time period when the soft, hard, and broad energy band images all had similar best-fit radii, beyond which the hard band images started becoming larger than their broadband counterparts (Figure \ref{Fig:radial expansion}b). Before 2012 (day $\sim$9000), the hard band images were smaller, likely because of a significant contribution from the low density H II region interior to the ER. Later ($\sim$2012-2016), even in the hard band, ER (inter-clump regions) emission dominates, and hence all sub-band images have roughly the same size. After $\sim$2016, the hard band images become larger, as the increasing contribution of the spectrally-hard X-ray emission from the shocked low-density CSM beyond the ER dominates the hardband X-ray emission. The blast wave starting to move out around $\sim$2012 (9000 days) is also consistent with our latest deep Chandra high-resolution gratings spectroscopic results, where a significant decline in the density profile of the shocked CSM was detected between days $\sim$8800 and $\sim$11,350 \citep{Ravi21}. We note that while the origin of X-ray and radio emission may be different, a similar re-acceleration of the shock wave in SNR 1987A was also reported in the radio band \citep{Cendes18}.

\subsection{Compact Remnant in SN 1987A}
The presence of a Pulsar Wind Nebula (PWN) was suggested in SNR 1987A, based on the NuSTAR observations, in which hard X-ray emission is detected at E $\sim$10--20 keV \citep{Greco21, Greco22}. We note that the third hard spectral component (thermal or non-thermal) that was invoked to describe the NuSTAR spectrum of SNR 1987A \citep{Greco21, Alp21} is not required to fit the observed X-ray spectrum of SNR 1987A in the Chandra bandpass. While the origin of this hard X-ray emission is debated \citep{Alp21}, high-resolution ALMA data \citep{Cigan19} and recent IR spectra with JWST \citep{Fransson24} show evidence of the reprocessed emission due to photoionization of the surrounding cold ejecta from the neutron star, supporting the PWN interpretation.

We do not yet detect any emission from the central object in 0.3--8.0 keV Chandra images. This is not surprising, as recent estimates of the H column density due to the presence of cold ejecta toward the center of SNR 1987A are of order $\ge$10$^{23}$ cm$^{-2}$ \citep{Esposito18, Page20, Greco21}. At such high column densities, the potential signatures of a compact object will be unlikely to be detected below 10 keV. In the future, as the central cool ejecta gas thins out and the absorption column density decreases, Chandra imaging will be crucial to directly identify the presence of the putative point source.

\section{Conclusions}\label{sec:5} 

In this work, we report the results from our imaging-spectroscopic study of SNR 1987A based on 47 epochs of Chandra observation, spanning the last 23 years. We re-processed the entire datasets of our Chandra monitoring observations of SNR 1987A for a self-consistent analysis. Here we summarize our main results as below.

\begin{itemize}[]
    \item The soft X-ray light curve declines linearly between 2016 and 2020 (by $\sim$4.5\,\% yr$^{-1}$), consistent with a physical picture where the forward shock is now propagating into the lower density CSM outside the ER. \\
    \vspace{-0.5cm}
    \item Since 2018, we detect for the first time in the Chandra spectrum the Fe K emission line feature at $E$ $\sim$6.7 keV. We find hints for asymmetric distribution of the Fe K line flux between the eastern and western halves of the SNR. This development could be either due to the strengthening of emission from an asymmetric distribution of CSM and/or the asymmetric reverse-shock developing in SNR 1987A. Future Chandra observations will be crucial to discriminate between these origin scenarios.
    \item The soft X-ray light curve since 2020 has stabilized once again at $\sim$7 $\times$ 10$^{-12}$ erg s$^{-1}$ cm$^{-2}$. These observations, coupled with the presence of Fe K line increasing in significance, could suggest the onset of contributions from the reverse-shocked ejecta. Comparisons with MHD simulations show that the evolution of the observed soft X-ray light curve is consistent with increasing contributions from reverse shock and outer ejecta interactions.
    \item Since 2016, the peak emission radii measured in our Chandra images are expanding at a similar rate as before 2016 (1830 $\pm$ 40 $^{-1}$ km s$^{-1}$), suggesting that the dominant X-ray emission observed by Chandra is still within the ER. On the other hand, while the thickness of the X-ray emitting ring was constant between 2004 and 2016, it has increased systematically (by $\sim$35\,\%) between 2016 and 2022. This suggests that the outer boundary of the shock front in SNR 1987A is interacting with less-dense CSM beyond the dense ER. Also, the 3.0--8.0 keV hard images are increasing at a faster rate (2970 $\pm$ 150 km s$^{-1}$), growing larger than the broadband images as of 2022, which indicates a significant contribution of the spectrally-hard X-ray emission from the shocked low-density CSM beyond the ER in the hardband X-ray image. 

\end{itemize}

\facilities{This paper employs a list of Chandra datasets, obtained by the Chandra X-ray Observatory, contained in \dataset[DOI: https://doi.org/10.25574/cdc.214]{https://doi.org/10.25574/cdc.214}.}\\

We thank the anonymous referee for their careful and constructive consideration of our work. This work was supported in part by NASA Chandra grants, G07-18060A, G08-19057X, G08-19043B, G09-20051B, and G02-23037X. SO and MM acknowledge financial contribution from the PRIN MUR ``Life, death and after-death of massive stars: reconstructing the path from the pre-supernova evolution to the supernova remnant"







\end{document}